\documentclass[journal,draftclsnofoot,onecolumn,12pt]{IEEEtran}
\usepackage{amsthm,amssymb,graphicx,multirow,amsmath,color,amsfonts}
\usepackage[update,prepend]{epstopdf}
\usepackage[noadjust]{cite}
\usepackage[latin1]{inputenc}
\usepackage{tikz}
\usepackage{bbm} 
\usepackage{pdfpages}
\usepackage{multirow}
\usepackage{comment}
\usepackage{mwe}
\usepackage{subfig}

\newcommand{\sbt}{\,\begin{picture}(-1,1)(-1,-3)\circle*{3}\end{picture}\ }

\def\nb0{{\mathbf{0}}}
\def\nb1{{\mathbf{1}}}

\newtheorem{lemma}{Lemma}
\newtheorem{thm}{Theorem}
\newtheorem{definition}{Definition}

\newtheorem{prop}{Proposition}
\newtheorem{cor}{Corollary}

\newtheorem{remark}{Remark}

\begin{document}
\pagenumbering{gobble}
\graphicspath{{./Figures/}}
\title{
On the 3-D Placement of Airborne Base Stations Using Tethered UAVs 
}

\author{
Mustafa A. Kishk, Ahmed Bader, and Mohamed-Slim Alouini
\thanks{Mustafa A. Kishk and Mohamed-Slim Alouini are with King Abdullah University of Science and Technology (KAUST), Thuwal 23955-6900, Saudi Arabia (e-mail: mustafa.kishk@kaust.edu.sa; slim.alouini@kaust.edu.sa).\newline Ahmed Bader is with Insyab Wireless Limited, Dubai 1961, United Arab Emirates (e-mail: ahmed@insyab.com).
} 

}

\maketitle

\begin{abstract}
One of the main challenges slowing the deployment of airborne base stations (BSs) using unmanned aerial vehicles (UAVs) is the limited on-board energy and flight time. One potential solution to such problem, is to provide the UAV with power supply through a tether that connects the UAV to the ground. In this paper, we study the optimal placement of tethered UAVs (TUAVs) to minimize the average path-loss between the TUAV and a receiver located on the ground. Given that the tether has a maximum length, and the launching point of the TUAV (the starting point of the tether) is placed on a rooftop, the TUAV is only allowed to hover within a specific {\em hovering region}. Beside the maximum tether length, this hovering region also depends on the heights of the buildings surrounding the rooftop, which requires the inclination angle of the tether not to be below a given minimum value, in order to avoid tangling and ensure safety.  We first formulate the optimization problem for such setup and provide some useful insights on its solution. Next, we derive upper and lower bounds for the optimal values of the tether length and inclination angle. We also propose a suboptimal closed-form solution for the tether length and its inclination angle that is based on maximizing the line-of-sight probability. Finally, we derive the probability distribution of the minimum inclination angle of the tether length. We show that its mean value varies depending on the environment from $10^\circ$ in suburban environments to $31^\circ$ in high rise urban environments. Our numerical results show that the derived upper and lower bounds on the optimal values of the tether length and inclination angle lead to tight suboptimal values of the average path-loss that are only $0-3$ dBs above the minimum value.
\end{abstract}
\section{Introduction} \label{sec:intro}
With the increasing number of use cases and applications of unmanned aerial vehicles (UAVs) in the past few years~\cite{8660516,8675384,8438489,zeng2019accessing,8579209}, developing communication systems tailored to serve such high altitude nodes was inevitable. This is evident by the existence of UAV-related work items in recent releases of 3GPP~\cite{muruganathan2018overview}. As a result of relying on the cellular network to provide the communication support for those airborne users, many research works were motivated to study the fundamental differences between serving terrestrial and aerial users, main challenges, and possible solutions~\cite{8692749,8470897,azariarxiv,8337920,8528463}.

The research in that area has later evolved to studying the potential of using UAVs as flying base stations (BSs) to serve terrestrial users~\cite{8437232,7470937,7932923,7470933}. This is motivated by the improved channel quality between the BS and the user when the BS is deployed at high altitude, due to the high probability of establishing a line-of-sight (LoS) channel. In addition, 
the mobility and relocation capability of the envisioned UAV-mounted BS would highly increase the flexibility of its deployment, which is perfect in areas with time-varying traffic demand spatial distribution~\cite{7744808}. Furthermore, due to its easy and quick deployment (plug and play), UAV can be used for emergency scenarios and disaster recovery to serve mobile users in recovering areas~\cite{8360023}. Existing literature has considered multiple aspects of the flying BS system, such as trajectory optimization~\cite{7888557,8320772,8570843,8247211}, optimal deployment for limited hovering time~\cite{8053918}, interference analysis~\cite{7967745}, and coexistence with device-to-device (D2D) networks~\cite{7412759}.

Unfortunately, as promising as flying BSs might seem, practical limitations prevented it from attracting similar attention from the industrial sector, with limited exceptions that will be discussed later in this section. These limitations include: (i) achievable UAV payload, (ii) flight time, and (iii) on-board available energy for processing and communication. In order to use the UAV as a flying BS, we need to equip it with antennas and multiple processing units. However, achievable payloads for currently available UAVs are very limited. This, in turns, disables some important features such as sectorization and antenna diversity. In order to increase the payload of a UAV, we need to provide it with a stable source of energy, which is not available in untethered UAVs. 

The limited flight time of the UAV is one of the main obstacles in the road towards realizing UAV-mounted BSs. The current state of the art can only achieve less than an hour of hovering time before battery depletion. Hence, the UAV needs to revisit a ground station every hour to recharge or change the battery and then fly back to its hovering location. This leaves the UAV coverage area temporarily out of service, which reduces the performance of the cellular network.  

A cellular BS is one of the most power consuming components of the cellular network. It requires power for data transmission, processing, and backhauling. While 4G cell-sites consume around 6 kilowatts, 5G cell-sites are expected to consume around 10-15 kilowatts. With the available limited battery on-board in current UAVs, providing the required power for communication and processing is a challenge. Looking back at the above three main practical limitations, we observe that the key solution to achieve a reliable, stable, and sustainable flying BS is ensuring a stable source of energy for the UAV~\cite{8255733,8648453}. Fortunately, this is provided in {\em tethered UAVs} (TUAVs)~\cite{TUAVmag2019}. 

With a stable power supply provided to the UAV through a tether connected to the ground station (GS), the TUAV can achieve much longer flight times, support heavier payload, and support the required power for on-board communication and processing. In addition to providing the TUAV with power supply, a wired data link is also extended through the tether. This enables a wired backhaul link between the flying BS and the GS, which solves another significant challenge in untethered UAV-mounted BSs, namely, wireless backhaul communication~\cite{8255764,7470932,cicek2018backhaul,8422376}. Due to its great potential to realize a flying BS, many companies around the world have started developing TUAVs, which are actually available for commercial use. In Table~\ref{tab:companies}, we summarize the specifications of TUAVs implemented by these companies. 
\begin{table}\caption{TUAV state of the art}
\centering
    \begin{tabular}{| {c} | {c} | {c} |}
        \hline
     \textbf{Company name}& \textbf{Maximum tether length} & \textbf{Flight time}\\ \hline
        Equinox Systems~\cite{equinox} & 150 m & 30 Days \\ \hline
        TDS~\cite{tds}                 & 120 m & unlimited  \\ \hline
        Aria Insights~\cite{aria}      & 120 m & multi-day operation \\ \hline
        Elistair~\cite{elistair}       & 80 m & 10+ hours \\ \hline        
    \end{tabular}
\label{tab:companies}
\end{table}

Note that TUAVs are different from, the more famous, {\em Helikites} in multiple aspects. Helikites, were recently adopted for deploying flying BSs at high altitudes~\cite{7470932}. Helikites are designed in a way that combines some properties of kites and helium balloons. Thanks to the intelligent aerodynamic design of Helikites, they do not need to be power supplied in order to stay in the air. However, they are connected to the ground through a tether, which restrains the Helikite and prevents it from floating away. Just like TUAVs, Helikites are capable of supporting relatively large payloads. However, there is some specific operation conditions that need to be satisfied to support such large payloads. For instance, a 15 m$^3$ Helikite needs at least 15 miles/hour wind speed to be able to support a payload of 12 Kgs. This operation conditions are relaxed when the size of the Helikite is increased. If the Helikite size is increased to 34 m$^3$, it is able to support a payload of upto 14 Kgs without the need to any wind. For heavier payloads, higher speed wind is required. Such wind speeds might not be available at lower altitudes, which means that successful operation requires either high altitude deployment or increasing the Helikite sizes~\cite{abs}. On the other hand, TUAVs are much lower in size (less than 3 m$^3$~\cite{equinox}), but their reachable altitude is restricted by the tether length. Another concern that comes to picture when using Helikites for providing cellular coverage, is the energy requirements of its payload. Beside antennas, the Helikite also carries processing units on-board, which need to be powered. Given that the Helikite's tether does not provide it with power supply, the only option to power the on-board payload is carrying a battery on the Helikite, which limits the endurance of the flying BS~\cite{abs}. We can conclude from this discussion that there is a fundamental difference between TUAVs and Helikites in terms of architecture, achievable altitudes, payloads, design challenges, and operating conditions. Hence, we believe that the use cases and applications that would benefit from TUAVs and Helikites are quite different, which means that the two technologies are complementary to each other with similar importance for future wireless networks. 

Recently, TUAVs were used by an American service provider in Puerto Rico to provide cellular coverage for the recovering areas after the hurricane Maria~\cite{sundaresan2018skylite}. In fact, the majority of the companies mentioned in Table~\ref{tab:companies} rely on a specific set of applications for promoting their TUAV products such as surveillance, broadcasting, video streaming for assessment of critical situations, search and rescue, and providing cellular coverage for emergency scenarios until the damaged cell towers are rebuilt. However, the recent technological advances and the current achievable specifications, see~\cite{equinox} for example, make TUAVs a very attractive solution for non-emergency related scenarios such as offloading terrestrial BSs in locations with high traffic demand, providing cellular coverage in remote and rural areas, and network densification~\cite{TUAVmag2019}.

In order to consider widely deploying TUAVs for cellular coverage enhancement, performance of such setup should be carefully studied first. In particular, analysis of such systems should be performed while taking into account the special characteristics of the TUAV such as the limited tether length and the safety considerations to avoid tangling the tether upon surrounding buildings. These characteristics lead to a set of problems that are fundamentally different from the problems considered in the literature of untethered UAVs. In this paper, we study the optimal placement of a TUAV in order to minimize the average path-loss at the target receiver. Compared to typical untethered UAVs, the optimal placement problem of the TUAV is fundamentally different due to the restrictions introduced by the maximum tether length. Assuming the GS is deployed on a rooftop of a building, the value of the maximum tether length constraints the deployment of the TUAV within a hemisphere, centered at the rooftop. In addition, to ensure safe operation, the optimal placement problem should consider preventing the tether from tangling upon surrounding buildings, which might lead to tether cutting, and hence, TUAV crashing. More details on the contributions of this paper will be provided later in this section.
\subsection{Related Work}
TUAVs' share of the existing literature is very limited. The only works that shed some light on such setup are~\cite{8644135,8432474}. In~\cite{8644135}, the authors proposed using TUAVs in post-disaster scenarios, due to the existence of a wired data link through the tether, to provide a backhaul link for the untethered UAVs. In particular, while the untethered drones are dedicated to providing cellular coverage for the recovering area, they use a free-space-optical (FSO) link with the TUAV for backhaul. In~\cite{8432474}, the authors studied the optimal trajectory of an untethered UAV serving two users, which are apart with distance $D$. However, when the product of the distance and the velocity of the UAV is much lower than the flight time of the UAV, the problem considered in this paper reduces to finding the optimal hovering location, instead of the optimal trajectory. This can be considered equivalent to the deployment problem of a TUAV if the TUAV is always located exactly above its GS, with the tether length extended to its maximum value. This assumption highly relies on the main objective of the TUAV deployment. The optimal tether length and inclination angle are actually one of the most important results that we will provide later in this paper. 

As stated earlier, we aim in this paper to study the optimal placement problem of a TUAV, given the location of the GS and the maximum tether length value. The existing works in literature that worked on UAV optimal placement problems focused solely on untethered UAVs. One of the most important works in that direction was provided in~\cite{6863654}. In this paper, the authors focused on optimizing the altitude of the UAV to maximize its coverage radius. In particular, they provided an expression for the average path-loss expression as a function of the altitude and the distance between the UAV's projection and the receiver. Using that expression, the radius of the area where the receivers obtain an average path-loss below a predefined threshold is maximized. Authors in~\cite{7918510}, considered similar setup with the objective of minimizing the transmit power of the UAV. In~\cite{7762053}, the authors tackled the optimal placement problem with the objective of minimizing the number of required UAVs. In~\cite{7486987}, the authors considered a system of multi-antenna UAVs used to provide coverage for a given area. The coverage probability, defined as the probability that the signal-to-interference-plus-noise (SINR) is above a predefined threshold, is derived. The altitudes of the UAVs is then optimized to maximize the coverage probability using circle packing theory. In~\cite{7510820}, the UAV optimal placement problem was considered with the objective of maximizing the number of covered users, where a user is covered if its perceived path-loss is below a specific value. The results showed that the UAV deployment is efficient, in terms of average number of covered users, in suburban and urban regions. However, the mean number of covered users decreases dramatically in high rise urban regions. In~\cite{8038869}, a system of UAVs is used to collect data from an Internet of Things (IoT) network. For that setup, the authors optimize the deployment of the UAVs as well as their mobility from one location to another, based on the activity of the IoT devices. Reducing the mobility of the UAVs as much as possible is of great importance for untethered UAVs, because it consumes most of the available energy on-board. This concern, however, does not come to picture when TUAVs are deployed, due to the existence of a stable power supply through the tether. 

It can be observed from the above discussion that the main design parameters of an untethered UAV-enabled communication system are (i) the UAV's altitude, (ii) the location of its projection on the ground, and (iii) its trajectory. These parameters are typically optimized to maximize the coverage of the UAV and minimize its energy consumption, in order to increase the flight time. On the other hand, a TUAV does not have concerns related to mobility minimization, as stated, due to the existence of a stable source of energy through the tether. However, unlike untethered UAVs, the altitude and projection of the TUAV are constrained with the maximum tether length. Hence, compared to untethered UAVs, the optimal placement problem of a TUAV has some fundamental differences, which is what we aim to study in this paper. The contributions of this paper are summarized next.
\subsection{Contributions}
Compared to the existing literature on placement optimization of UAV-mounted BSs, which solely focused on untethered UAVs, this paper's  main objective is to optimize the location of a tethered UAV (TUAV) using its tether length and inclination angle as the main optimization parameters.  We consider a TUAV system where the GS (the starting point of the tether) is placed on the rooftop of a building and a receiver located at a given distance from the building. Aligning with available TUAV state of the art, we assume that the tether has a maximum length, which limits its mobility and relocation capability. In addition, motivated by the importance of the tether for supporting the TUAV with both power and data, we assume that the inclination angle of the tether can not be below a specific value. This value ensures that the tether will not get tangled upon any of the surrounding builds.
\subsubsection{A Novel Mathematical Framework for Modeling and Analysis of TUAVs} 
We provide a mathematical model for the achievable locations by the TUAV in the 3-dimensional (3-D) plane, which we refer to as the {\em hovering region}. Next, we formalize the optimal placement problem that aims to minimize the average path-loss at the receiver, with the hovering region as the main constraint of the optimization problem. Before solving the problem, we carefully analyze it and provide multiple insights on its solution.
\subsubsection{Upper and Lower Bounds for the Optimal Values of the Tether Length and Inclination Angle}
Using the drawn insights on the placement problem, we derive upper and lower bounds for the optimal values of the optimal tether length and inclination angle. In particular, when the distance between the receiver and the building is below a certain threshold, we show that the optimal value of the inclination angle is its minimum value and provide upper and lower bounds for the tether length. When the distance is above the threshold, we show that the optimal tether length is its maximum value and derive upper and lower bounds for the optimal inclination angle. We show using numerical results, assuming a dense urban environment, that the upper and lower bounds lead to tight results with respect to the optimal average path-loss value.
\subsubsection{Closed-Form Expressions for Suboptimal Solution}
We propose a suboptimal solution for the TUAV placement problem that is based on maximizing the probability of LoS. We derive closed form expressions for the suboptimal values of the tether length, inclination angle, and average path-loss. We evaluate the tightness of this suboptimal solution in a dense urban environment using numerical results. The results show that the suboptimal value of the average path-loss is only $0-3$ dBs above the optimal value.
\subsubsection{Probability Distribution of the Minimum Inclination Angle in Different Environments}
As stated earlier, the minimum allowed value for the inclination angle is based on the altitude of the rooftop with respect to its surrounding buildings. The inclination angle should be high enough to prevent tangling. Based on that approach, we use tools from stochastic geometry to model the locations of the buildings and concretely derive the probability distribution of the inclination angle minimum value. We show that the inclination angle's mean value varies from $10^\circ$ at suburban environments to $31^\circ$ at high rise urban regions. 
\section{System Setup}
\begin{figure}
    \centering
\includegraphics[width=0.7\columnwidth]{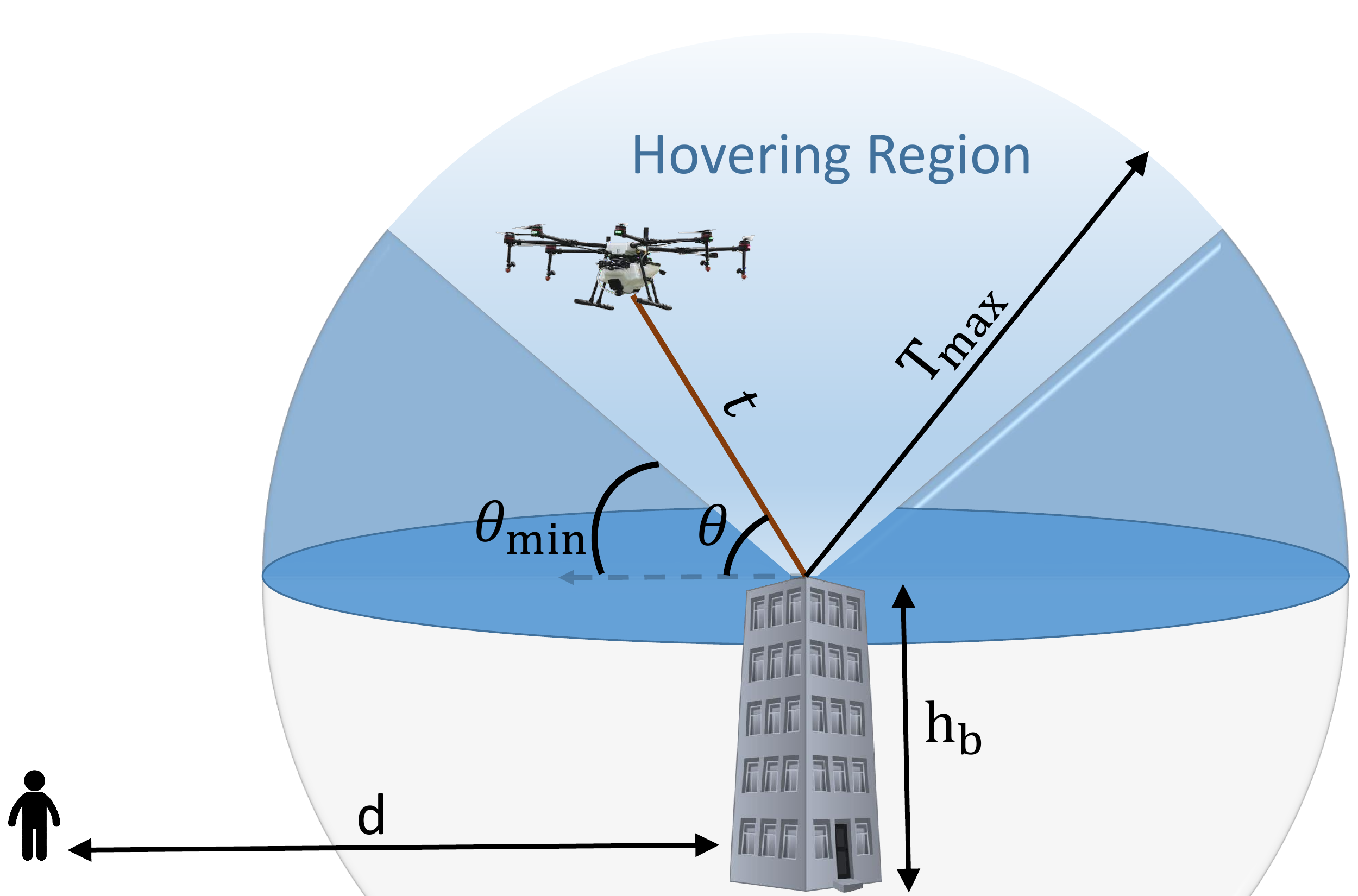}
\caption{The system setup considered in this paper.}
\label{fig:sys1}
\end{figure}

As shown in Fig.~\ref{fig:sys1}, we consider a system composed of a TUAV launched from a ground station (GS) that is placed on a rooftop at height $h_b$. The TUAV has the freedom to hover anywhere within the hemisphere centered at the rooftop with radius equal to the maximum value of the tether length $T_{\rm max}$. In order to avoid tangling upon surrounding buildings and ensure safety of the tether, the inclination angle $\theta$ of the tether, has a minimum value $\theta_{\rm min}$, as shown in Fig.~\ref{fig:sys1}. We use Cartesian coordinates $(x,y,z)$ in the rest of the paper, to represent the locations of the TUAV and the receiver. Without loss of generality, we assume that the receiver is located at the origin $(0,0,0)$, while the rooftop is located at the point $(d,0,h_b)$, where $d$ is the distance between the receiver and the building. Based on values of $\theta_{\rm min}$, $h_b$, and $T_{\rm max}$, the TUAV can be deployed anywhere within a specific {\em hovering region}, which is defined next.
\begin{definition}
The hovering region of the TUAV, is the set of locations in $\mathbb{R}^3$ that are reachable by the TUAV:
\begin{align}
\mathcal{M}=\Bigg\{(x,y,z):&\sqrt{(x-d)^2+(y)^2+(z-h_b)^2}<=T_{\rm max},\nonumber\\ &{\rm sin}^{-1}\left(\frac{z_p-h_b}{\sqrt{(x_p-d)^2+(y_p)^2+(z_p-h_b)^2}}\right)\geq \theta_{\rm min}\Bigg\}.
\end{align}
\end{definition}
The TUAV aims to find the location within the hovering region that has the minimum average path-loss (PL). Given that the TUAV is placed at a location $p=(x_p,y_p,z_p)$, the average PL is defined next.
\begin{definition}
The average PL between a TUAV located at $p$ and the receiver is~\cite{6863654}:
\begin{align}\label{eqn:1}
{\rm PL_p}=P_{\rm LoS}(p)R_p^2\eta_{\rm LoS}+(1-P_{\rm LoS}(p))R_p^2\eta_{\rm nLoS},
\end{align}
where $P_{\rm LoS}$ is the line-of-sight (LoS) probability, $R_p=\sqrt{x_p^2+y_p^2+z_p^2}$ is the distance between the receiver and the TUAV, and $\eta_{\rm LoS}<\eta_{\rm nLoS}$ are the mean excessive path-loss values for the cases of LoS and non-LoS, respectively. 
\end{definition}
Next, we formally define the LoS probability, which highly impacts the value of PL, and, in turns, the optimal placement problem.
\begin{definition}
The LoS probability between a TUAV located at $p$ and the receiver is
\begin{align}\label{eqn:plos}
P_{\rm LoS}(p)=a\left({\rm tan}^{-1}\left(\frac{z_p}{\sqrt{x_p^2+y_p^2}}\right)-15\right)^{b},
\end{align}
where the values of the parameters $a$ and $b$ depend on whether the environment is subruban, urban, dense urban, or high rise urban~\cite{7037248}. 
\end{definition}
The ultimate goal of TUAV placement is to find the location, within its hovering region, that minimizes the average PL between the TUAV and the receiver. The optimization problem can be formally written as follows. 
\begin{subequations}
\begin{alignat}{2}
&{\rm {\mathbf{OP_1}:}}\ \ \ \underset{p\in\mathbb{R}^3}{\rm minimize}        \ \ \ \ {\rm PL}_{p}\nonumber\\
&\ \ \ \ \ \ \ \ \ \ \ \text{subject to:} &      & \nonumber\\
&\ \ \ \ \ \ \ \ \ \ t_p=\sqrt{(x_p-d)^2+(y_p)^2+(z_p-h_b)^2}<=T_{\rm max}, &      &\label{eq:constraint1}\\
&\ \ \ \ \ \ \ \ \ \  \theta_p={\rm sin}^{-1}\left(\frac{z_p-h_b}{\sqrt{(x_p-d)^2+(y_p)^2+(z_p-h_b)^2}}\right)\geq \theta_{\rm min},&      &\label{eq:constraint2}
\end{alignat}
\end{subequations}
where constraint (\ref{eq:constraint1}) ensures that the tether length is below its maximum value and constraint (\ref{eq:constraint2}) ensures that the tether inclincation angle is above its minimum value.
\begin{prop}\label{prop:pre}
For any two points $p_1,p_2\in\mathbb{R}^3$, if $z_{p_1}=z_{p_2}$ and $R_{p_1}<R_{p_2}$, then ${\rm PL}_{p_1}<{\rm PL}_{p_2}$.
\begin{IEEEproof}
Given that $z_{p_1}=z_{p_2}$ and $R_{p_1}<R_{p_2}$, we know that ${\sqrt{x_{p_1}^2+y_{p_1}^2}}<{\sqrt{x_{p_2}^2+y_{p_2}^2}}$. Revisiting (\ref{eqn:plos}), we conclude that $P_{\rm LoS}(p_1)>P_{\rm LoS}(p_2)$. Given that $\eta_{\rm LoS}<\eta_{\rm nLoS}$, the statement in the proposition follows.
\end{IEEEproof}
\end{prop}
\begin{table}[t]\caption{Table of notations}
\centering
\begin{center}
\resizebox{\textwidth}{!}{
\renewcommand{\arraystretch}{1.4}
    \begin{tabular}{ {c} | {c} }
    \hline
        \hline
    \textbf{Notation} & \textbf{Description} \\ \hline
    ${\rm PL}_p$ & The average value of the path-loss between a TUAV located at $p\in\mathbb{R}^3$ and the receiver located at the origin \\ \hline
    $R_p$ & The distance between a TUAV located at $p\in\mathbb{R}^3$ and the receiver  \\ \hline
        $P_{\rm LoS}(p)$ & The LoS probability between a TUAV located at $p\in\mathbb{R}^3$ and the receiver  \\ \hline
        $t_p$ & The length of the tether when the TUAV is placed at $p\in\mathbb{R}^3$\\ \hline
        $\theta_p$ & The inclination angle of the tether with respect to the x-y plane when the TUAV is placed at $p\in\mathbb{R}^3$\\ \hline
        $\theta_{\rm min}$ & The minimum allowed inclination angle, to avoid tangling with surrounding buildings\\ \hline
        $\eta_{\rm LoS}$; $\eta_{\rm nLoS}$ &the mean value of the excessive path-loss for the cases of LoS and non-LoS, respectively \\ \hline
        $t_{\rm opt}$; $t_{\rm sub}$ & Optimal; suboptimal values of the tether length\\ \hline
        $\theta_{\rm opt}$; $\theta_{\rm sub}$ & Optimal; suboptimal values of the tether inclination angle\\ \hline
    \end{tabular}}
\end{center}
\label{tab:TableOfNotations}
\end{table}
\begin{prop}\label{prop:y}
The optimal location for the TUAV, $p_{\rm opt}$, satisfies the following:
\begin{align}
y_{p_{\rm opt}}=0.
\end{align}
\begin{IEEEproof}
For any point, $p$, with $|y_p|>0$, that satisfies (\ref{eq:constraint1}) and (\ref{eq:constraint2}), its projection on the x-z plane $\{y=0\}$, $\hat{p}$, has the following characteristics:
\begin{align}
z_{\hat{p}}\overset{(a)}{=}z_{p},\ x_{\hat{p}}=x_{p},\ R_{\hat{p}}\overset{(b)}{<}R_{p}.\nonumber\label{eqn:prop:y}
\end{align}
Hence, given that $|y_{\hat{p}}|<|y_p|$ and $p$ satisfies constraints $(\ref{eq:constraint1})$ and $(\ref{eq:constraint2})$, then $\hat{p}$ also satisfies constraints $(\ref{eq:constraint1})$ and $(\ref{eq:constraint2})$. Also, based on $(a)$, and $(b)$, and using Proposition~\ref{prop:pre}, we conclude that ${\rm PL}_{\hat{p}}<{\rm PL}_{p}$.

\end{IEEEproof}
\end{prop}

\begin{prop}\label{prop:x1}
The optimal location for the TUAV, $p_{\rm opt}$, satisfies the following:
\begin{align}
x_{p_{\rm opt}}<d.
\end{align}
\begin{IEEEproof}
For any point, $p$, with $x_p>d$, that satisfies (\ref{eq:constraint1}) and (\ref{eq:constraint2}), consider a point $\hat{p}$ with the following characteristics:
\begin{align}
z_{\hat{p}}{=}z_{p},\ x_{\hat{p}}=2d-x_{p},\ y_{\hat{p}}=y_{p}.\label{eqn:prop:x1}
\end{align}
Hence, given that $(x_{\hat{p}}-d)^2=(x_{p}-d)^2$, then $\hat{p}$ also satisfies constraints (\ref{eq:constraint1}) and (\ref{eq:constraint2}). In addition, since $x_p>d$, then $x_{\hat{p}}^2<x_p^2$. Hence, $R_{\hat{p}}<R_{p}$ and $P_{\rm LoS}(\hat{p})>P_{\rm LoS}(p)$, which implies that ${\rm PL_{\hat{p}}}<{\rm PL_{p}}$. This concludes the proof.
\end{IEEEproof}
\end{prop}
\begin{figure}
    \centering
\includegraphics[width=1\columnwidth]{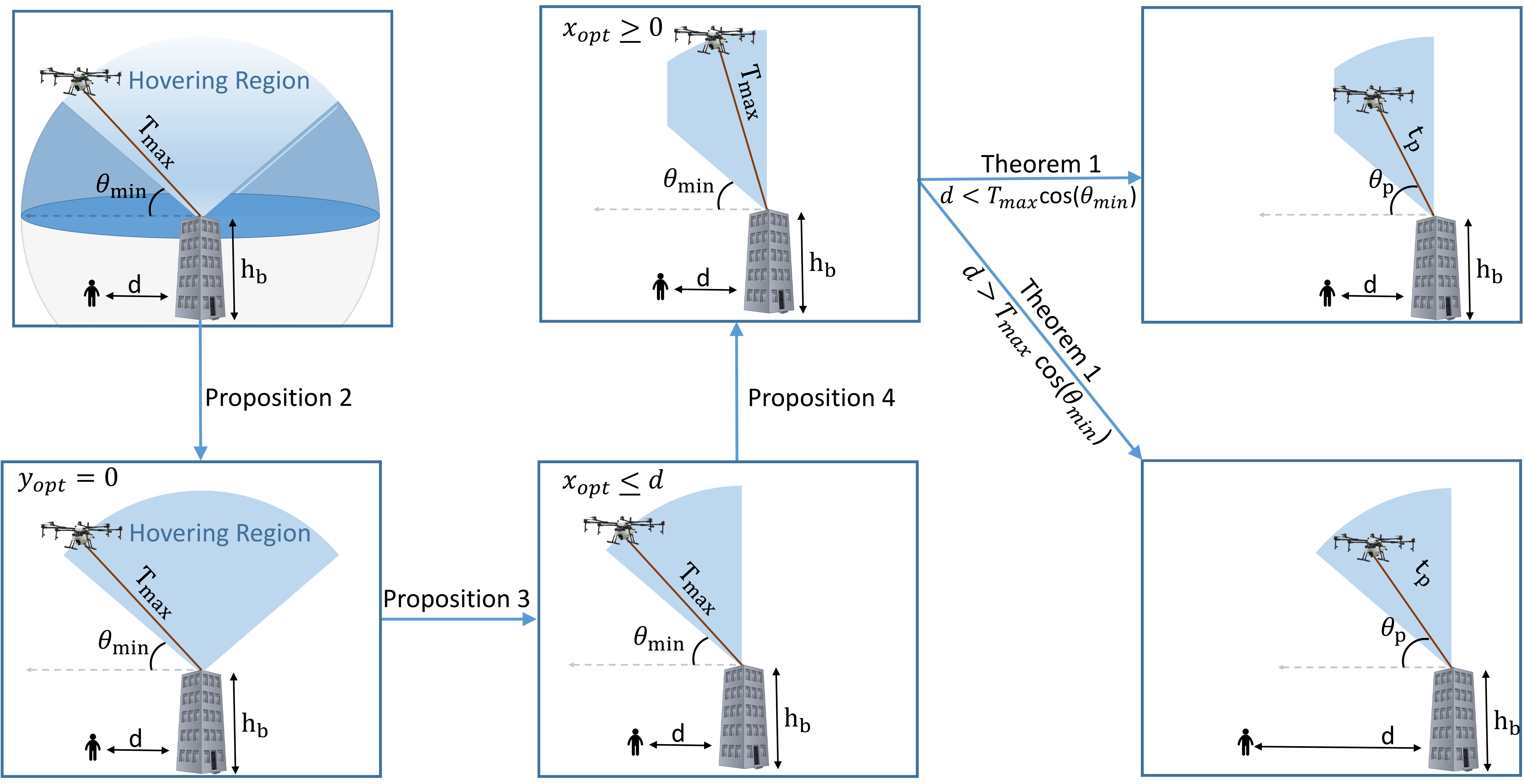}
\caption{The effect of Propositions~\ref{prop:y},~\ref{prop:x1}, and~\ref{prop:x2} on the hovering region.}
\label{fig:1}
\end{figure}
\begin{prop}\label{prop:x2}
The optimal location for the TUAV, $p_{\rm opt}$, satisfies the following:
\begin{align}
x_{p_{\rm opt}}\geq 0.
\end{align}
\begin{IEEEproof}
For any point, $p$, with $x_p<0$, that satisfies (\ref{eq:constraint1}) and (\ref{eq:constraint2}), consider a point $\hat{p}$ with the following characteristics:
\begin{align}
z_{\hat{p}}{=}z_{p},\ x_{\hat{p}}=0,\ y_{\hat{p}}=y_{p}.
\end{align}
Hence, given that $x_p<0$, then $(x_{\hat{p}}-d)^2<(x_{p}-d)^2$, which ensures that $\hat{p}$ also satisfies constraints (\ref{eq:constraint1}) and (\ref{eq:constraint2}). Also, since $x_{\hat{p}}^2<x_p^2$, we conclude that $R_{\hat{p}}<R_{p}$ and $P_{\rm LoS}(\hat{p})>P_{\rm LoS}(p)$, which implies that ${\rm PL_{\hat{p}}}<{\rm PL_{p}}$.
\end{IEEEproof}
\end{prop}
Given that $t$ is the length of tether and $\theta$ is the angle between the tether and the $x-y$ plane, we can use the above propositions and the fact that $x=d-t{\rm cos}(\theta)$ and $z=h_b+t{\rm sin}(\theta)$ to rewrite the optimization problem ${\rm \mathbf{OP_1}}$ in terms of $t$ and $\theta$ instead of $x$, $y$, and $z$. The equivalent optimization problem is provided in the following theorem.
\begin{thm}\label{thm:op2}
The optimal TUAV placement problem ${\rm \mathbf{OP_1}}$ is equivalent to ${\rm \mathbf{OP_2}}$, which is presented next.
\begin{subequations}
\begin{alignat}{2}
&{\rm {\mathbf{OP_2}:}}\ \ \ \underset{t_p,\theta_p}{\rm minimize}        \ \ \ \ {\rm PL}_{p}\nonumber\\
&\ \ \ \ \ \ \ \ \ \ \ \text{subject to:} &      & \nonumber\\
&\ \ \ \ \ \ \ \ \ \ \theta_{\rm min}\leq\theta_p\leq\frac{\pi}{2}, &      &\label{eq:constraint21}\\
&\ \ \ \ \ \ \ \ \ \  t_p{\rm cos}(\theta_p)\leq d,&      &\label{eq:constraint22}\\
&\ \ \ \ \ \ \ \ \ \  0\leq t_p\leq T_{\rm max},&      &\label{eq:constraint23}
\end{alignat}
\end{subequations}
where constraints (\ref{eq:constraint21}) and (\ref{eq:constraint22}) ensure that $0<x<d$ as discussed in Propositions~\ref{prop:x1} and~\ref{prop:x2}, and constraint (\ref{eq:constraint23}) ensures that the tether length is less than its maximum value.
\end{thm}
\begin{remark}
It can be observed from Theorem~\ref{thm:op2} that having $d>T_{\rm max}{\rm cos}(\theta_{\rm min})$ implies that $d>t_{p}{\rm cos}(\theta_p)$ for all $t_p<T_{\rm max}$ and $\theta_{\rm min}\leq\theta_p\leq\frac{\pi}{2}$. In other words, if $d>T_{\rm max}{\rm cos}(\theta_{\rm min})$, constraint (\ref{eq:constraint22}) is always satisfied. This case is shown in Fig.~\ref{fig:1} for more clarification.
\end{remark}
\section{Upper and Lower Bounds on the Optimal Solution}
\begin{figure}
    \centering
\includegraphics[width=0.7\columnwidth]{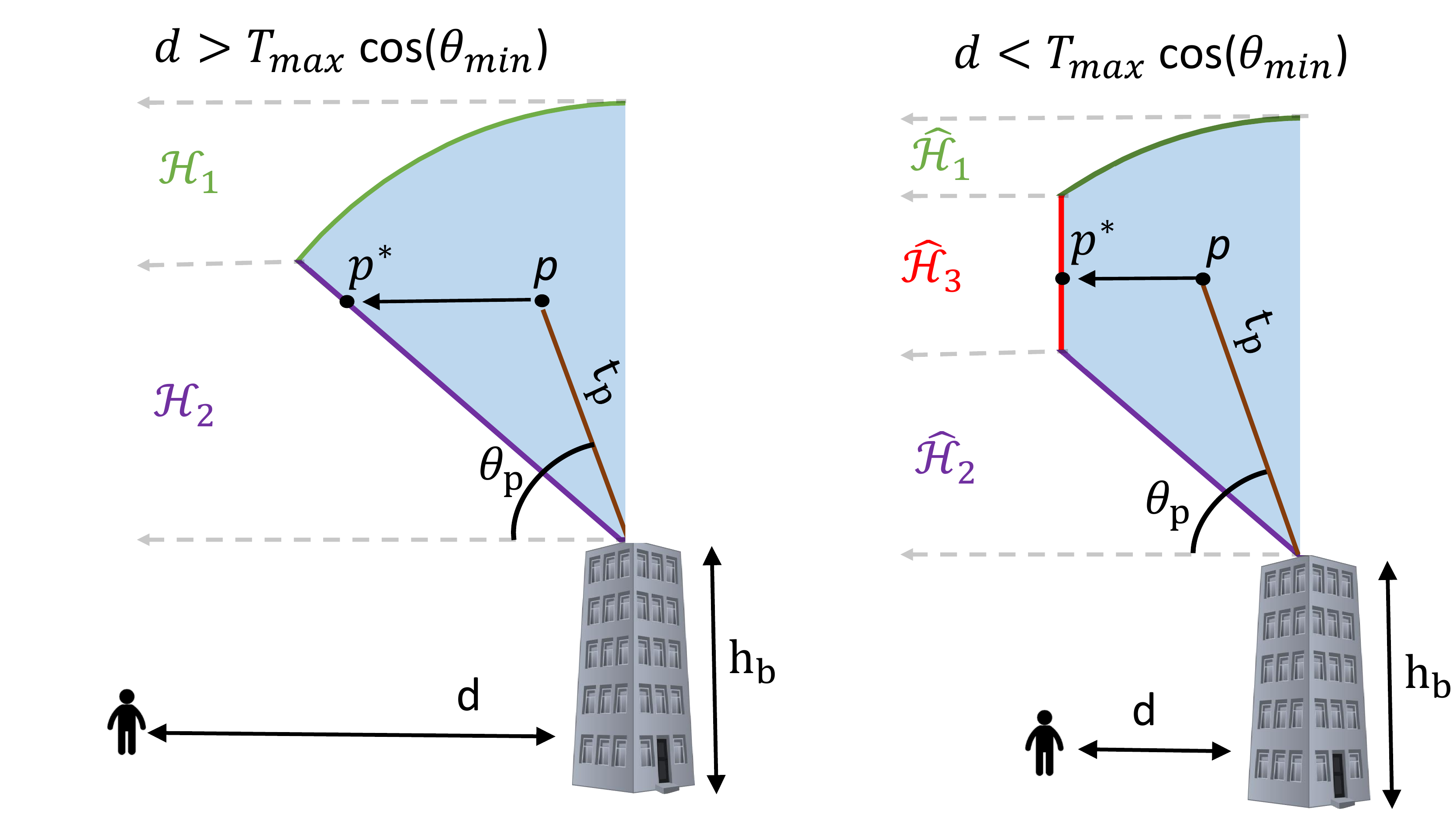}
\caption{As explained in Theorem~\ref{thm2}, ${\rm PL}_{p}$ is always greater than ${\rm PL}_{p^*}$.}
\label{fig:thm2}
\end{figure}
In this section, we aim to study the optimization problem in Theorem~\ref{thm:op2} and provide upper and lower bounds for the optimal values of the tether length $t_p$ and the inclination angle $\theta_p$. In the following theorem, we show that the optimal location of the TUAV lies on the border of the hovering region, between the rooftop and the receiver, as shown in Fig.~\ref{fig:thm2}.
\begin{thm}\label{thm2}
When $d>T_{\rm max}{\rm cos}(\theta_{\rm min})$, the optimal values of $(t_p,\theta_p)$ belong to the following set
\begin{align}
\mathcal{H}=\mathcal{H}_1\cup\mathcal{H}_2,
\end{align}
where 
\begin{align}
\mathcal{H}_1=\left\{(t,\theta):\theta_{\rm min}\leq\theta\leq\frac{\pi}{2},t=T_{\rm max}\right\},
\end{align}
and
\begin{align}
\mathcal{H}_2=\left\{(t,\theta):\theta=\theta_{\rm min},0<t<T_{\rm max}\right\}.
\end{align}
When $d<T_{\rm max}{\rm cos}(\theta_{\rm min})$, the optimal values of $(t_p,\theta_p)$ belong to the following set
\begin{align}
\hat{\mathcal{H}}=\hat{\mathcal{H}}_1\cup\hat{\mathcal{H}}_2\cup\hat{\mathcal{H}}_3,
\end{align}
where 
\begin{align}
\hat{\mathcal{H}}_1&=\left\{(t,\theta):{\rm cos}^{-1}\left(\frac{d}{T_{\rm max}}\right)\leq\theta\leq\frac{\pi}{2},t=T_{\rm max}\right\},\\
\hat{\mathcal{H}}_2&=\left\{(t,\theta):\theta=\theta_{\rm min},0<t\leq\frac{d}{{\rm cos}(\theta_{\rm min})}\right\},\\
\hat{\mathcal{H}}_3&=\left\{(t,\theta):\theta_{\rm min}\leq\theta\leq{\rm cos}^{-1}\left(\frac{d}{T_{\rm max}}\right),t=\frac{d}{{\rm cos}(\theta)}\right\}.
\end{align}
\begin{IEEEproof}
To avoid repetition, we will focus in the proof on the case of $d>T_{\rm max}{\rm cos}(\theta_{\rm min})$. Recalling that for any point $p$, satisfying the three constraints in ${\rm \mathbf{OP_2}}$, we know that $z_p=h_b+t_p{\rm sin}(\theta_p)$ and $x_p=h_b-t_p{\rm cos}(\theta_p)$. Now, as shown in Fig.~\ref{fig:thm2}, for any point $p\not\in\mathcal{H}$, its projection $p^*$ on $\mathcal{H}$ has the following characteristics:\\
\begin{equation}
\label{muli_def}
z_{p^*}=\left\{
	\begin{array}{ll}
		h_b+T_{\rm max}{\rm sin}(\theta_{p^*})  & \mbox{if } {p^*}\in \mathcal{H}_1 \\
		h_b+t_{p^*}{\rm sin}(\theta_{\rm min}) & \mbox{if } {p^*} \in \mathcal{H}_2
	\end{array}.
\right.
\end{equation}
Given that $z_{p^*}=z_{p}$, $\theta_p>\theta_{\rm min}$, and $t_p<T_{\rm max}$, then we can conclude that $\theta_{p^*}<\theta_p$ and $t_p<t_{p^*}$. Hence, it can be easily shown that $x_p>x_{p^*}$, which means that $R_p>R_{p^*}$. Using Proposition~\ref{prop:pre}, we can show that ${\rm PL}_{p}>{\rm PL}_{p^*}$, which concludes the proof.
\end{IEEEproof}
\end{thm}
\begin{cor}\label{cor1}
For the case of $d<T_{\rm max}{\rm cos}(\theta_{\rm min})$, the optimal values of $(t_p,\theta_p)$ belong to $\hat{\mathcal{H}}_2=\{(t,\theta):\theta=\theta_{\rm min},0<t\leq\frac{d}{{\rm cos}(\theta_{\rm min})}\}$.
\begin{IEEEproof}
The point $p=\left(t_p=\frac{d}{{\rm cos}(\theta_{\rm min})},\theta_p=\theta_{\rm min}\right)\in\hat{\mathcal{H}}_2$ has an elevation angle $\tan^{-1}\left(\frac{z_p}{\sqrt{y_p^2+x_p^2}}\right)=\tan^{-1}\left(\frac{h_b+t_p\sin (\theta_p)}{d-t_p\cos (\theta_p)}\right)=\frac{\pi}{2}$ and $R_p=h_b+d\tan (\theta_{\rm min})$. Hence, recalling~(\ref{eqn:plos}), for any point $p^*\in\hat{\mathcal{H}}_1\cup\hat{\mathcal{H}}_3$, we have $P_{\rm LoS}(p)\geq P_{\rm LoS}(p^*)$ and $R_{p}<R_{p^*}$, which means that ${\rm PL}_{p}<{\rm PL}_{p^*}$.
\end{IEEEproof}
\end{cor}
Now, in the following two lemmas, we provide some important insights on $R_p$ and $P_{\rm LoS}(p)$ that will be useful for defining upper and lower bounds on the optimal values of $t_p$ and $\theta_p$.
\begin{lemma}\label{lemma1}
For any given $\theta_p$, where $\theta_{\rm min}<\theta_p<\frac{\pi}{2}$, $R_p$ is a convex function of $t_p$ where ${\rm arg\ min}\underset{t_p}{\ }R_p=t^*(\theta_p)=d\cos (\theta_p)-h_b\sin (\theta_p)$. In addition, $P_{\rm LoS}(p)$ is an increasing function of $t_p$. 
\begin{IEEEproof}
See Appendix~\ref{app:lemma1}.
\end{IEEEproof}
\end{lemma}
\begin{lemma}\label{lemma2}
For any given $t_p$, where $0<t_p\leq T_{\rm max}$, $R_p$ is an increasing function of $\theta_p$ in the set $\mathcal{S}=\{\theta_{\rm min}<\theta_p<\frac{\pi}{2}\}$. In addition, $P_{\rm LoS}(p)$ is a concave function of $\theta_p$ in the set $\mathcal{S}$, where
\begin{align}
{\rm arg\ max}\underset{\theta_p}{\ }P_{\rm LoS}(p)=\theta^*(t_p)=\sin^{-1}\left(\frac{d}{\sqrt{d^2+h^2}}\right)-\sin^{-1}\left(\frac{t_{p}}{\sqrt{d^2+h^2}}\right).
\end{align}
\begin{IEEEproof}
See Appendix~\ref{app:lemma2}.
\end{IEEEproof}
\end{lemma}
In the following theorem, we provide upper and lower bounds for the optimal values of $t$ and $\theta$ for different values of $d$.
\begin{thm}\label{thm3}
The solution to ${\rm \mathbf{OP_2}}$, provided in Theorem~\ref{thm:op2} is $(t_p,\theta_p)\in\mathcal{H}_{\rm opt}$ where
\begin{equation}
\label{muli_def2}
\mathcal{H}_{\rm opt}=\left\{
	\begin{array}{ll}
		\{(t_{\rm opt},\theta_{\rm min}):\ \max(0,t^*(\theta_{\rm min}))\leq t_{\rm opt}\leq \frac{d}{\cos(\theta_{\rm min})}\},  & \mbox{if } {d}\leq T_{\rm max}\cos (\theta_{\rm min}) \\
		\{(t_{\rm opt},\theta_{\rm min}):\ \max(0,t^*(\theta_{\rm min}))\leq t_{\rm opt}\leq T_{\rm max}\},  & \mbox{if } T_{\rm max}\cos (\theta_{\rm min})\leq {d}\leq \mathcal{F}\\
		\{(T_{\rm max},\theta_{\rm opt}):\ \theta_{\rm min}\leq \theta_{\rm opt}\leq \theta^*(T_{\rm max})\},  & \mbox{if } {d}\geq\mathcal{F}
	\end{array},
\right.
\end{equation}
${\rm max}(k,m)$ is the maximum value between $k$ and $m$, $\mathcal{F}=\frac{T_{\rm max}}{\cos (\theta_{\rm min})}+h_b\tan (\theta_{\rm min})$, $t^*(\theta_{\rm min})=d\cos (\theta_{\rm min})-h_b\sin (\theta_{\rm min})$, and $\theta^*(T_{\rm max})=\sin^{-1}\left(\frac{d}{\sqrt{d^2+h^2}}\right)-\sin^{-1}\left(\frac{T_{\rm max}}{\sqrt{d^2+h^2}}\right)$.
\begin{IEEEproof}
We start with the case of $d\leq T_{\rm max}\cos(\theta_{\rm min})$. In that case, as shown in Corollary~\ref{cor1}, the optimal value of $\theta$ is $\theta_{\rm min}$, and the optimal value of $t$ falls in the range $0<t\leq\frac{d}{\cos(\theta_{\rm min})}$. Recalling Lemma~\ref{lemma1}, and given that PL$_p$ is an increasing function of $P_{\rm LoS}(p)$ and a decreasing function of $R_p$, we can easily conclude that $t^*(\theta_{\rm min})\leq t_{\rm opt}$.\\
For the case of $T_{\rm max}\cos(\theta_{\rm min})\leq d\leq \mathcal{F}$, it can easily be shown that $t^*(\theta_{\rm min})<T_{\rm max}$ and $\theta^*(T_{\rm max})<\theta_{\rm min}$. Hence, recalling Theorem~\ref{thm2}, we can show that $(\theta_{\rm min},T_{\rm max})$ is the optimal point in $\mathcal{H}_1$ by using Lemma~\ref{lemma2}. Similarly, revisiting $\mathcal{H}_2$ in Theorem~\ref{thm2}, we can also show that $t_{\rm opt}\geq t^*(\theta_{\rm min})$, using Lemma~\ref{lemma1}.\\
For the case of $d\geq \mathcal{F}$, it can be shown that $t^*(\theta_{\rm min})>T_{\rm max}$ and $\theta^*(T_{\rm max})>\theta_{\rm min}$. Hence, we can show that $(\theta_{\rm min},T_{\rm max})$ is the optimal point in $\mathcal{H}_2$ by using Lemma~\ref{lemma1}. In addition, we can use Lemma~\ref{lemma2} to show that $\theta_{\rm opt}\leq\theta^*(T_{\rm max})$. This concludes the proof. 

\end{IEEEproof}
\end{thm}
\begin{cor}\label{cor2}
For the case of $\theta_{\rm min}=0^\circ$, the results in Theorem~\ref{thm3} reduce to
\begin{equation}
\label{muli_def222}
(t_{\rm opt},\theta_{\rm opt})=\left\{
	\begin{array}{ll}
		(d,0^\circ),\ & \mbox{if } {d}\leq T_{\rm max} \\
		(T_{\rm max},\theta_{\rm opt}),\ 0^\circ\leq \theta_{\rm opt}\leq \theta^*(T_{\rm max}),  & \mbox{if } {d}\geq T_{\rm max}
	\end{array}.
\right.
\end{equation}
\end{cor}
\begin{figure}
    \centering
\includegraphics[width=1\columnwidth]{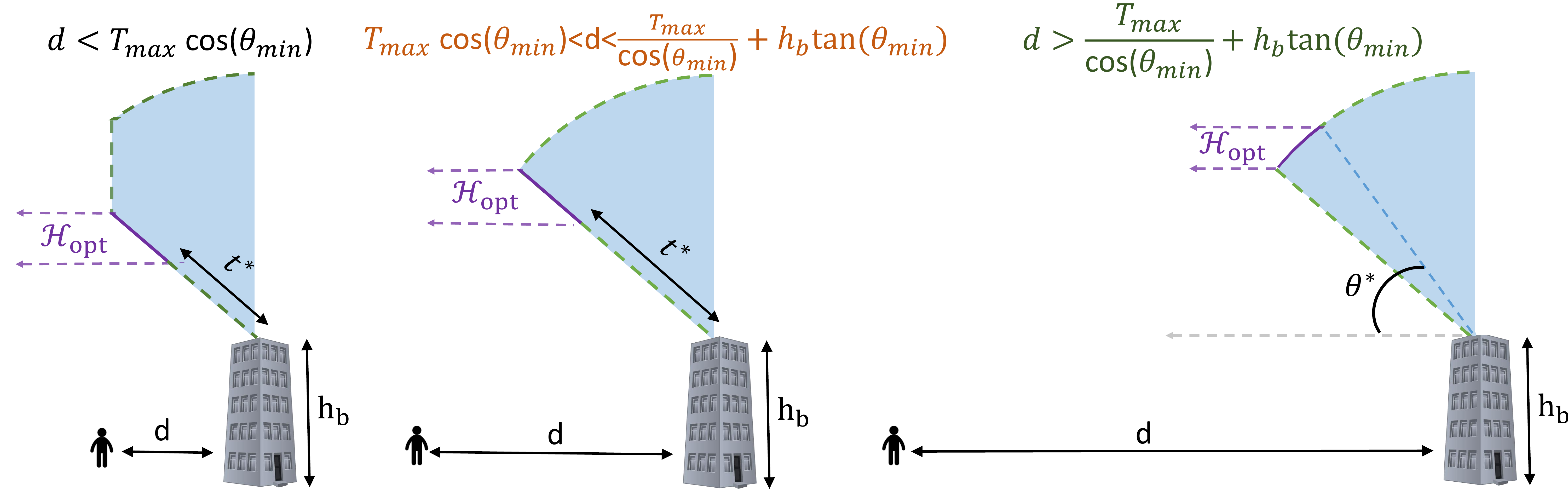}
\caption{The regions we should search for the optimal location of the TUAV $\mathcal{H}_{\rm opt}$, for different values of $d$.}
\label{fig:thm3}
\end{figure}
\begin{remark}\label{rem2}
In Fig.~\ref{fig:thm3}, we show how the results in Theorem~\ref{thm3} highly reduce the search range $\mathcal{H}_{\rm opt}$ for the optimal values of $(t_{\rm opt},\theta_{opt})$. However, in order to evaluate the efficiency of tightening the search range into $\mathcal{H}_{\rm opt}$, we need to evaluate the values of PL for all $(t_p,\theta_p)\in\mathcal{H}_{\rm opt}$, which is presented later in the results and discussion section. 
\end{remark}
\section{Suboptimal Solution}
In this section, we propose a suboptimal solution to \textbf{OP$_2$} that is based on maximizing the probability of LoS ($P_{\rm LoS}$) instead of minimizing the average path-loss (${\rm PL}$). Applying the comments in Lemmas~\ref{lemma1} and~\ref{lemma2} to find the point $p$ within the bounds defined in Theorem~\ref{thm3} that maximizes the value of $P_{\rm LoS}$, we define the suboptimal values $(t_{\rm sub},\theta_{\rm sub})$ as follows
\begin{equation}
(t_{\rm sub},\theta_{\rm sub})=\left\{
	\begin{array}{ll}
		(\frac{d}{\cos(\theta_{\rm min})},\theta_{\rm min}),\  & \mbox{if } {d}\leq T_{\rm max}\cos (\theta_{\rm min}) \\
		(T_{\rm max},\theta_{\rm min}),\  & \mbox{if } T_{\rm max}\cos (\theta_{\rm min})\leq {d}\leq \mathcal{F}\\
		(T_{\rm max},\theta^*(T_{\rm max})),\ & \mbox{if } {d}\geq\mathcal{F}
	\end{array},
\right.
\end{equation}
where $\mathcal{F}=\frac{T_{\rm max}}{\cos (\theta_{\rm min})}+h_b\tan (\theta_{\rm min})$ and $\theta^*(T_{\rm max})=\sin^{-1}\left(\frac{d}{\sqrt{d^2+h^2}}\right)-\sin^{-1}\left(\frac{T_{\rm max}}{\sqrt{d^2+h^2}}\right)$. In the following theorem, we provide the suboptimal values of ${\rm PL}$.
\begin{thm}\label{thm4}
The suboptimal value of ${\rm PL}$ when the TUAV is placed at the location that maximizes $P_{\rm LoS}$ is
\begin{align}
{\rm PL}^{\rm sub}=P_{\rm LoS}^{\rm sub}(R^{\rm sub})^2\eta_{\rm LoS}+(1-P_{\rm LoS}^{\rm sub})(R^{\rm sub})^2\eta_{\rm nLoS},
\end{align}
where
\begin{equation}
\label{muli_def3}
P_{\rm LoS}^{\rm sub}=\left\{
	\begin{array}{ll}
		a\left(90-15\right)^{b},\  & \mbox{if } {d}\leq T_{\rm max}\cos (\theta_{\rm min}) \\
		a\left({\rm tan}^{-1}\left(\frac{h_b+T_{\rm max}\sin(\theta_{\rm min})}{d-T_{\rm max}\cos(\theta_{\rm min})}\right)-15\right)^{b},\  & \mbox{if } T_{\rm max}\cos (\theta_{\rm min})\leq {d}\leq \mathcal{F}\\
		a\left({\rm tan}^{-1}\left(\frac{h_b\sqrt{h_b^2+d^2-T_{\rm max}^2}+dT_{\rm max}}{d\sqrt{h_b^2+d^2-T_{\rm max}^2}-h_bT_{\rm max}}\right)-15\right)^{b},\ & \mbox{if } {d}\geq\mathcal{F}
	\end{array},
\right.
\end{equation}
and
\begin{equation}
\label{muli_def4}
R^{\rm sub}=\left\{
	\begin{array}{ll}
		h_b+d\tan(\theta_{\rm min}),\  & \mbox{if } {d}\leq T_{\rm max}\cos (\theta_{\rm min}) \\
		\sqrt{h_b^2+d^2+T_{\rm max}^2-2dT_{\rm max}\cos(\theta_{\rm min})+2h_bT_{\rm max}\sin(\theta_{\rm min})},\  & \mbox{if } T_{\rm max}\cos (\theta_{\rm min})\leq {d}\\ &\leq \mathcal{F}\\
		\sqrt{h_b^2+d^2-T_{\rm max}^2},\ & \mbox{if } {d}\geq\mathcal{F}
	\end{array}.
\right.
\end{equation}
\begin{IEEEproof}
The above results follow directly by substituting for $x_p=d-t_{\rm sub}\cos(\theta_{\rm sub})$, $y_p=0$, and $z_p=h_b+t_{\rm sub}\sin(\theta_{\rm sub})$ in (\ref{eqn:1}) and (\ref{eqn:plos}).
\end{IEEEproof}
\end{thm}
\begin{cor}\label{cor3}
For the case of $\theta_{\rm min}=0^\circ$, the results in Theorem~\ref{thm4} reduce to
\begin{equation}
\label{muli_def333}
P_{\rm LoS}^{\rm sub}=\left\{
	\begin{array}{ll}
		a\left(90-15\right)^{b},\  & \mbox{if } {d}\leq T_{\rm max} \\
		a\left({\rm tan}^{-1}\left(\frac{h_b\sqrt{h_b^2+d^2-T_{\rm max}^2}+dT_{\rm max}}{d\sqrt{h_b^2+d^2-T_{\rm max}^2}-h_bT_{\rm max}}\right)-15\right)^{b},\ & \mbox{if } {d}\geq T_{\rm max}
	\end{array},
\right.
\end{equation}
and
\begin{equation}
\label{muli_def444}
R^{\rm sub}=\left\{
	\begin{array}{ll}
		h_b,\  & \mbox{if } {d}\leq T_{\rm max} \\
		\sqrt{h_b^2+d^2-T_{\rm max}^2},\ & \mbox{if } {d}\geq T_{\rm max}
	\end{array}.
\right.
\end{equation}
\end{cor} 
As it can be observed from Theorems~\ref{thm3} and~\ref{thm4}, the value of $\theta_{\rm min}$ highly impacts the optimal placement of the TUAV due to inducing a minimum value on the inclination angle of the tether. Hence, in the next section, we derive the distribution of $\theta_{\rm min}$.
\section{The distribution of $\theta_{\rm min}$}
The value of $\theta_{\rm min}$ depends on the height of the rooftop ($h_b$) at which the TUAV's launching point is placed, as well as the heights of the surrounding buildings. The computation of its value highly depends on the safety regulations followed in the deployment area. In this section, we assume that if there exists a building at a distance $L<T_{\rm max}$ that has a height $h>h_b$ then the minimum inclination angle of the tether is $\theta_{\rm min}=\tan^{-1}\left(\frac{h-h_b}{L}\right)$. This assumption prevents tether tangling upon this building and ensures its safety. In order to compute the distribution of $\theta_{\rm min}$, we model the locations of the surrounding buildings as a Poisson point process (PPP) $\Phi_b=\{x_i\}\in\mathbb{R}^2$ with density $\beta$ building/km$^2$. In addition, similar to~\cite{6863654}, the height of each building is assumed to be Rayleigh distributed with mean $\gamma$ meters. The values of $\beta$ and $\gamma$ vary depending on the environment, as discussed in~\cite{6863654}, as follows
  \begin{equation}
(\beta,\gamma)=\left\{
	\begin{array}{ll}
		(750,8),\  & \text{\rm for suburban environments} \\
		(500,15),\  & \text{\rm for urban environments} \\
		(300,20),\  & \text{\rm for dense urban environments} \\
		(300,50),\  & \text{\rm for high rise urban environments} 				
	\end{array}.
\right.
\end{equation}
\begin{figure}
    \centering
\includegraphics[width=0.5\columnwidth]{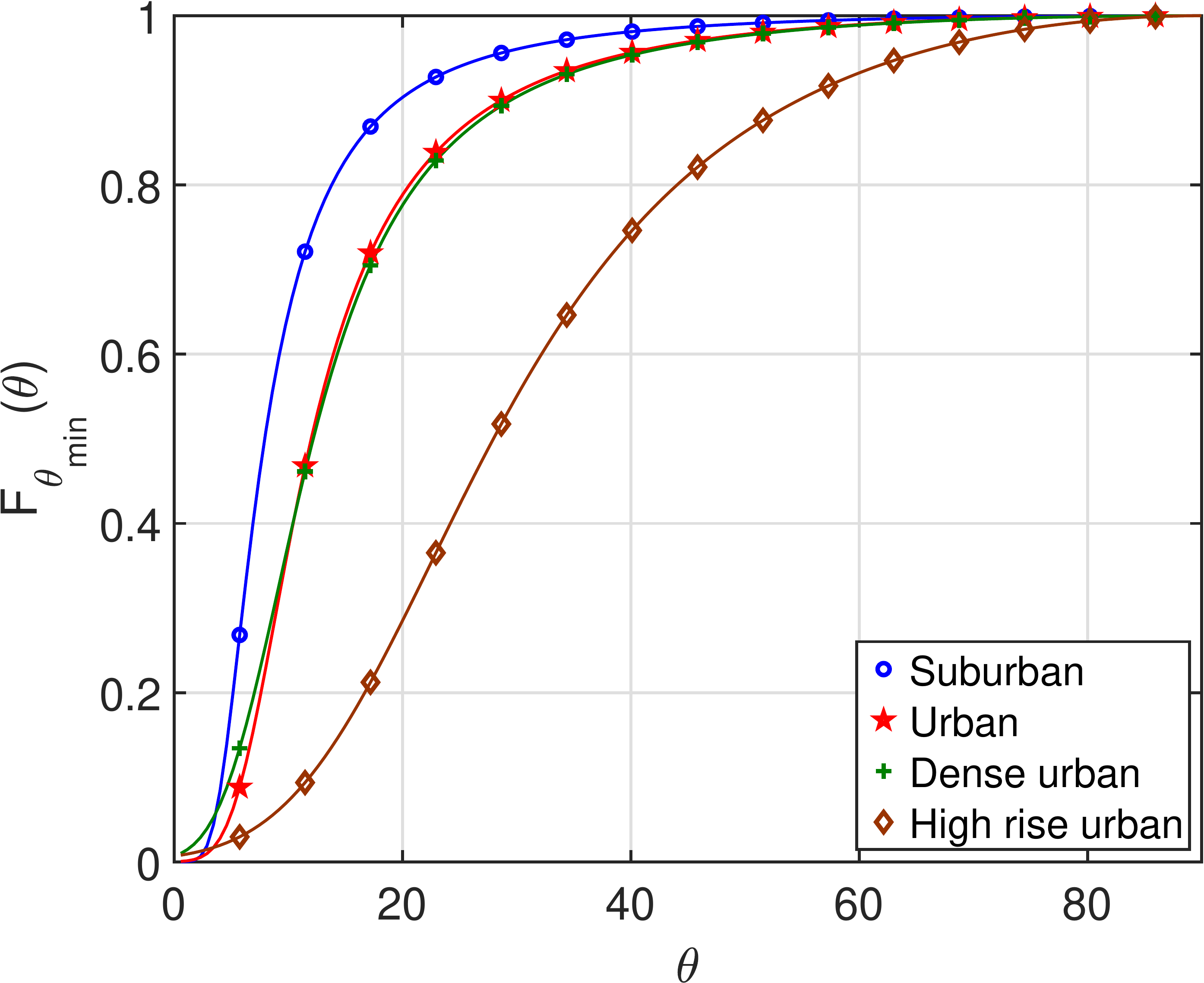}
\caption{The CDF of $\theta_{\rm min}$ for different environments.}
\label{fig:6}
\end{figure}
In the following theorem, we present the CDF of $\theta_{\rm min}$.
\begin{thm}\label{thm5}
The CDF of of $\theta_{\rm min}$ is
\begin{align}
F_{\theta_{\rm min}}(\theta)=\mathbb{P}(\theta_{\rm min}\leq\theta)=\exp\Bigg(\frac{-\pi\beta\gamma}{\tan^2(\theta)}\Bigg(&\gamma\left[\exp\left(-\frac{h_b^2}{\gamma^2}\right)-\exp\left(-\frac{(h_b+T_{\rm max}\sin(\theta))^2)}{\gamma^2}\right)\right]\nonumber\\&-h_b\left[\Gamma\left(\frac{1}{2},\frac{(h_b+T_{\rm max}\sin(\theta))^2)}{\gamma^2}\right)-\Gamma(\frac{1}{2},\frac{h_b^2}{\gamma^2})\right]\Bigg)\Bigg),
\end{align}
where $\Gamma({\sbt},{\sbt})$ is the lower incomplete Gamma function.
\begin{IEEEproof}
See Appendix~\ref{app:theta}.
\end{IEEEproof}
\end{thm}
\begin{remark}
As described in detail in Appendix~\ref{app:theta}, the event $\theta_{\rm min}<\theta$ takes place when every building at distance $L$, for any $L\leq T_{\rm max}\cos(\theta)$, has a height less than $h_b+L\tan(\theta)$. In other words, we only care for the heights of the buildings inside the ball $\mathcal{B}\big(0,T_{\rm max}\cos(\theta)\big)$. Hence, an interesting trade-off between the value of $T_{\rm max}$ and the CDF of $\theta_{\rm min}$ can be observed. In particular, at lower values of $T_{\rm max}$, the area of $\mathcal{B}\big(0,T_{\rm max}\cos(\theta)\big)$ is small. Hence, $\mathbb{P}(\theta_{\rm min}<\theta)$ is relatively high. However, as we increase the value of $T_{\rm max}$, the area of $\mathcal{B}\big(0,T_{\rm max}\cos(\theta)\big)$ increases, and hence, the probability decreases. The effect of this trade-off on the size of the hovering region should be carefully investigated, which is considered an interesting extension for our work.
\end{remark}
We plot the derived distribution of $\theta_{\rm min}$ in Fig.~\ref{fig:6} for different environments while assuming that $h_b=\gamma$. Agreeing with intuition, the CDF decreases as we move from suburban to high rise urban environments. The average value of $\theta_{\rm min}$ can be easily computed using the CDF: $\mathbb{E}[\theta_{\rm min}]=\int_0^{\frac{\pi}{2}}1-F_{\theta_{\rm min}}(\theta){\rm d}\theta$, which leads to the following values:
  \begin{equation}
\mathbb{E}[\theta_{\rm min}]=\left\{
	\begin{array}{ll}
		10.6^\circ,\  & \text{\rm for suburban environments} \\
		15.3^\circ,\  & \text{\rm for urban environments} \\
		15.3^\circ,\  & \text{\rm for dense urban environments} \\
		30.8^\circ,\  & \text{\rm for high rise urban environments} 				
	\end{array}.
\right.
\end{equation}
\section{Numerical Results and Discussion}
\begin{table}[]\caption{System parameters}
\centering
    \begin{tabular}{ |{c} | {c} | {c} | {c}|}
        \hline
    \textbf{Parameter} & \textbf{Value} & \textbf{Parameter} & \textbf{Value} \\ \hline
$T_{\rm max}$         & 150 m & $h_b$         & 30 m      \\ \hline
$\theta_{\rm min}$         & $0^{\circ}$  & Environment & Dense urban      \\ \hline
a          & 0.37    & b          & 0.21      \\ \hline
$\eta_{\rm LoS}  $        & $1.6$ dBs  & $\eta_{\rm nLoS}$  & $23$ dBs             \\ \hline
    \end{tabular}
\label{tab:parameters}
\end{table}
\begin{figure}
    \centering
\includegraphics[width=1\columnwidth]{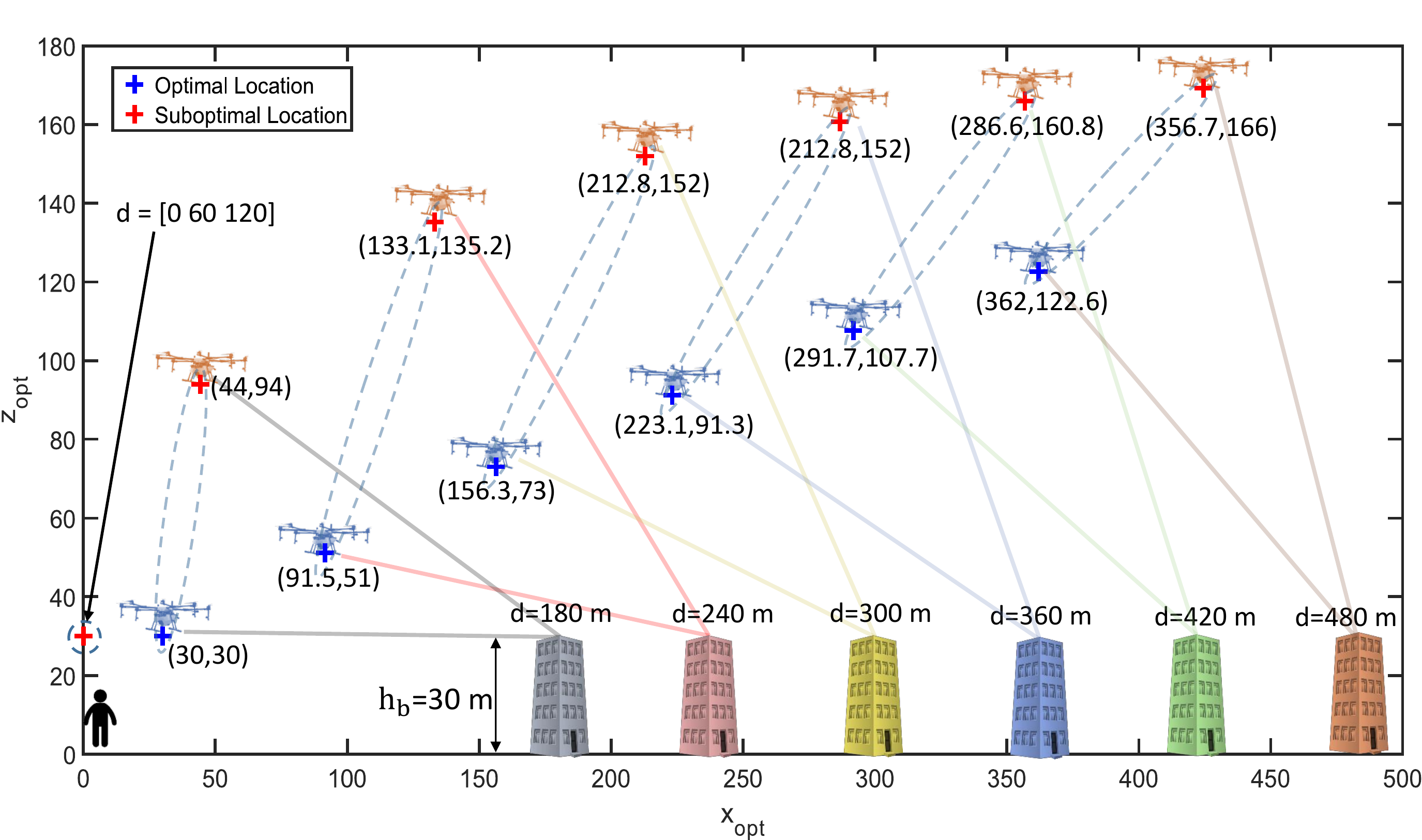}
\caption{The optimal and suboptimal locations of the TUAV as we increase the value of $d$.}
\label{fig:2}
\end{figure}
In this section, we evaluate the results provided in this paper. In particular, we evaluate the tightness of the upper and lower bounds provided in Theorem~\ref{thm3}, as well as the suboptimal solution provided in Theorem~\ref{thm4}. The values of the system parameters used in this section, unless otherwise is specified, are summarized in Table~\ref{tab:parameters}. In Fig.~\ref{fig:2}, we show the optimal and suboptimal locations of the TUAV as we increase the value of $d$. We notice that as long as $d<T_{\rm max}$, both optimal and suboptimal locations of the TUAV are similar, which perfectly supports our results in Corollary~\ref{cor2}. As the value of $d$ increases, we notice that the suboptimal locations of the TUAV have significantly higher altitudes compared to the optimal locations. This behavior results from the approach used to compute the suboptimal locations, which is maximizing the LoS probability $P_{\rm LoS}$. Given that $\theta_{\rm min}=0^\circ$, the values of $t_{\rm opt}$ and $\theta_{\rm opt}$ should be within the bounds provided in Corollary~\ref{cor2}, which can be easily verified using the values of $(x_{\rm opt},z_{\rm opt})$ provided in Fig.~\ref{fig:2} and the transformations $\theta_{\rm opt}=\tan^{-1}\left(\frac{z_{\rm opt}-h_b}{d-x_{\rm opt}}\right)$ and $t_{\rm opt}=\sqrt{(z-h_b)^2+(d-x_{\rm opt})^2}$.
\begin{figure}
\begin{minipage}{1\linewidth}
    \centering    
\includegraphics[width=0.5\columnwidth]{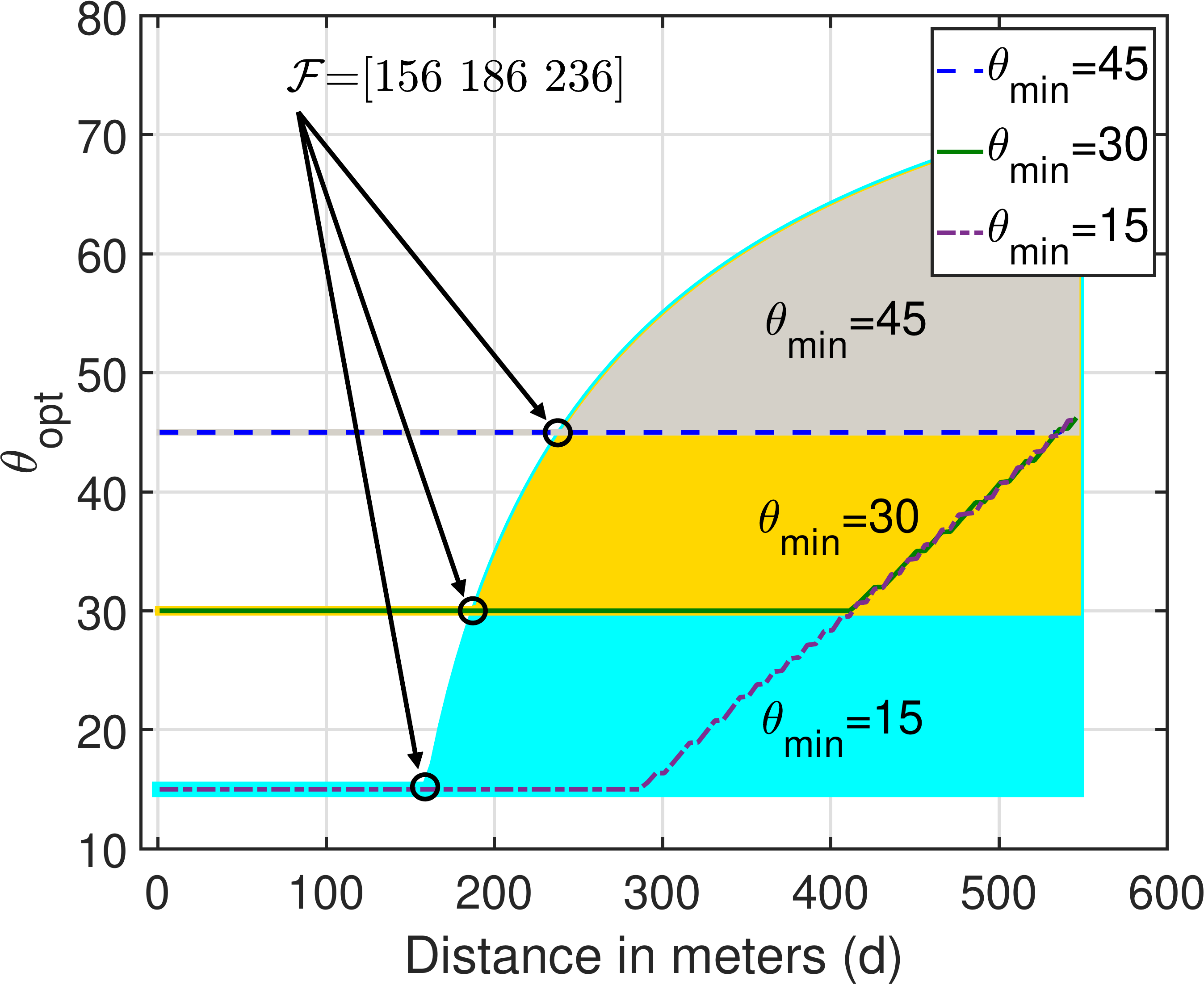}
\caption{The optimal value of $\theta$ is always within the bounds derived in Theorem~\ref{thm3}.}
\label{fig:77}
\hfill
\hfill
\end{minipage}
\begin{minipage}{1\linewidth}
    \centering
\includegraphics[width=0.5\columnwidth]{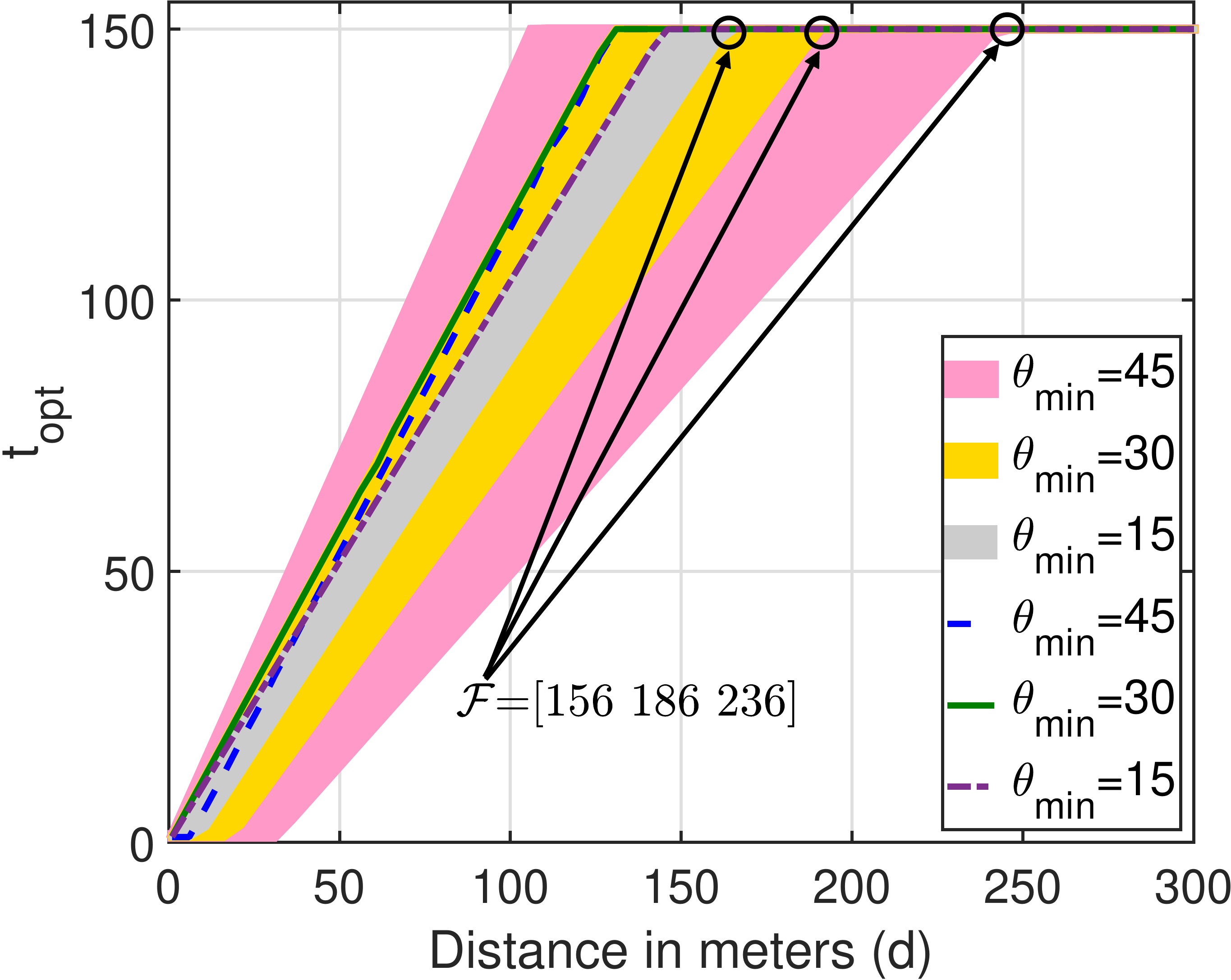}
\caption{The optimal value of $t$ is always within the bounds derived in Theorem~\ref{thm3}.}
\label{fig:88}
\end{minipage}
\end{figure}

In Theorem~\ref{thm3}, we provided upper and lower bounds for the optimal values of the tether length and inclination angle $(t_{\rm opt},\theta_{\rm opt})$. To verify these bounds, we provide the values of $\theta_{\rm opt}$ and $t_{\rm opt}$ for different values of $\theta_{\rm min}$ in Figures~\ref{fig:77} and~\ref{fig:88}. In each figure, we also show, in the colored area, the range between the upper and the lower bounds provided in Theorem~\ref{thm3}. The results prove that $(\theta_{\rm opt},t_{\rm opt})\in\mathcal{H}_{\rm opt}$, which is provided in Theorem~\ref{thm3}. We also verify that $\theta_{\rm opt}=\theta_{\rm min}$ as long as $d<\mathcal{F}$, and $t_{\rm opt}=T_{\rm max}$ as long as $d>\mathcal{F}$.

\begin{figure}
\begin{minipage}{1\linewidth}
    \centering    
\includegraphics[width=0.5\columnwidth]{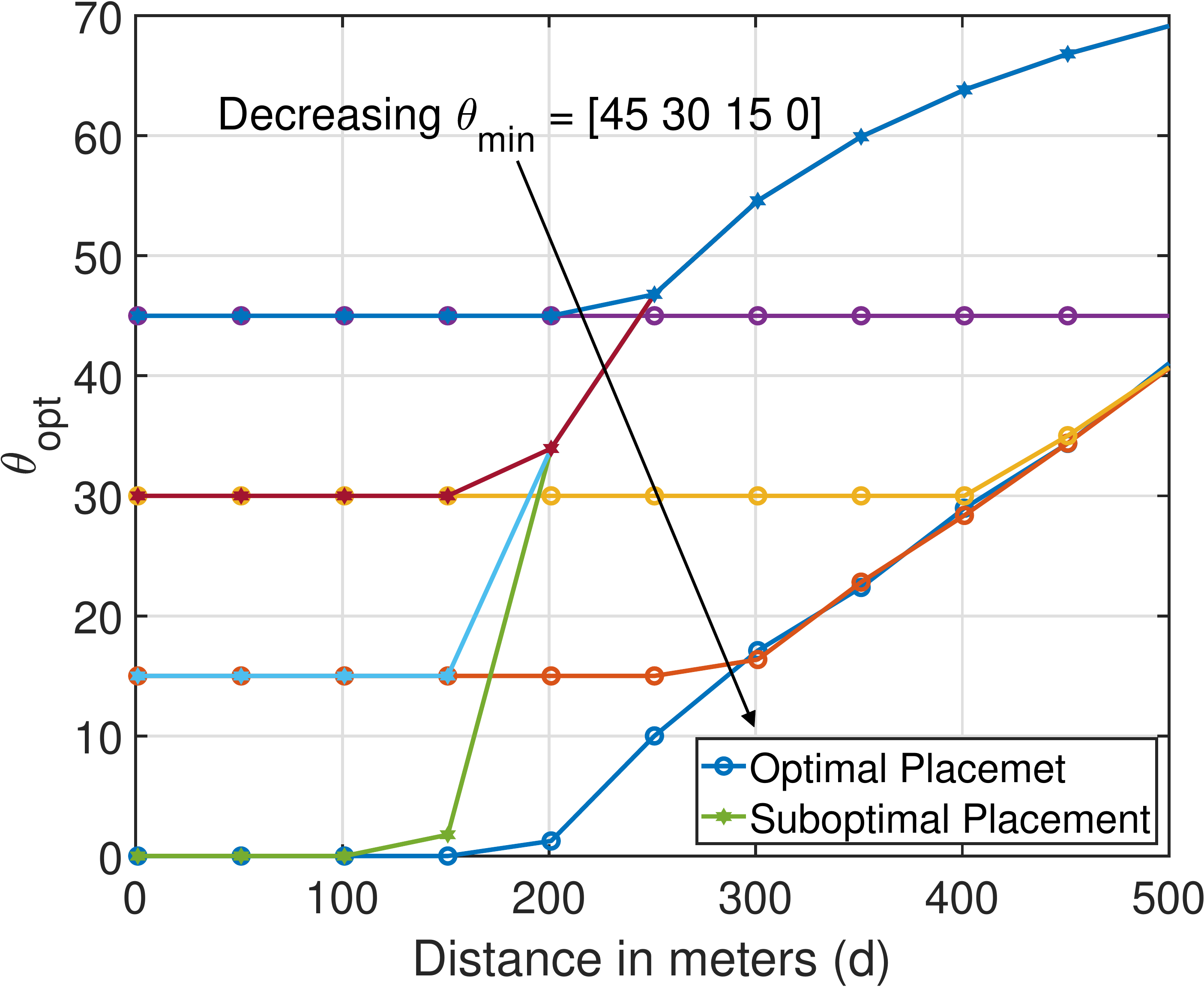}
\caption{Optimal and suboptimal values of $\theta$ for different values of $d$.}
\label{fig:99}
\hfill
\end{minipage}
\begin{minipage}{1\linewidth}
    \centering    
\includegraphics[width=0.5\columnwidth]{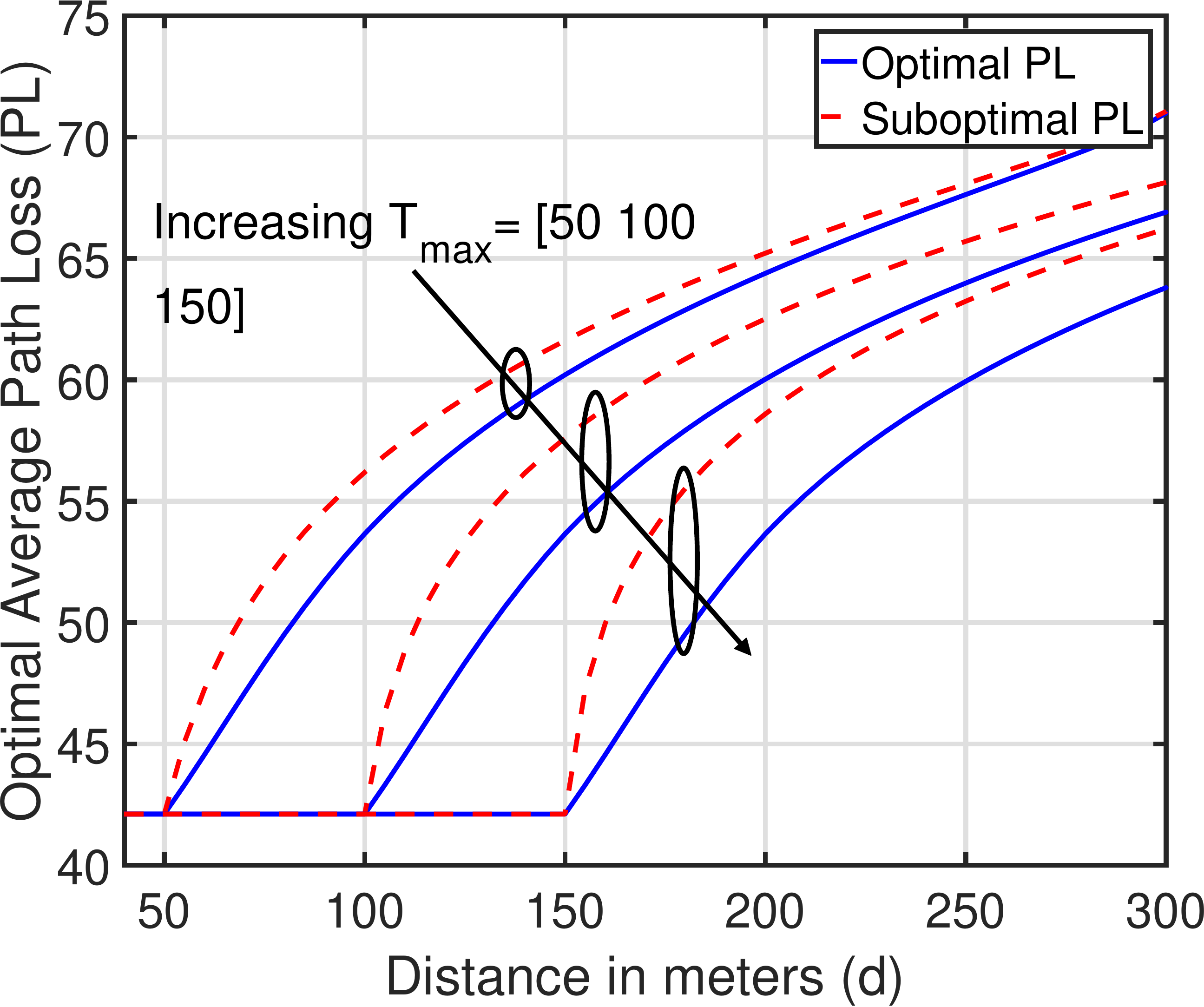}
\caption{Optimal and suboptimal values of PL for different values of $d$ and $T_{\rm max}$.}
\label{fig:4}
\end{minipage}
\end{figure}

In Fig.~\ref{fig:99}, we compare the optimal values of $\theta_{\rm opt}$ to the suboptimal values provided in Theorem~\ref{thm4}. As expected, since the suboptimal solution aims to maximize the value of $P_{\rm LoS}$, we can observe that $\theta_{\rm sub}>\theta_{\rm opt}$. However, to evaluate the tightness of the proposed suboptimal solution, we have to compare the values of PL, which is discussed next.

\begin{figure}
\begin{minipage}{.5\linewidth}
\centering
\subfloat[]{\label{fig:3a}\includegraphics[scale=.27]{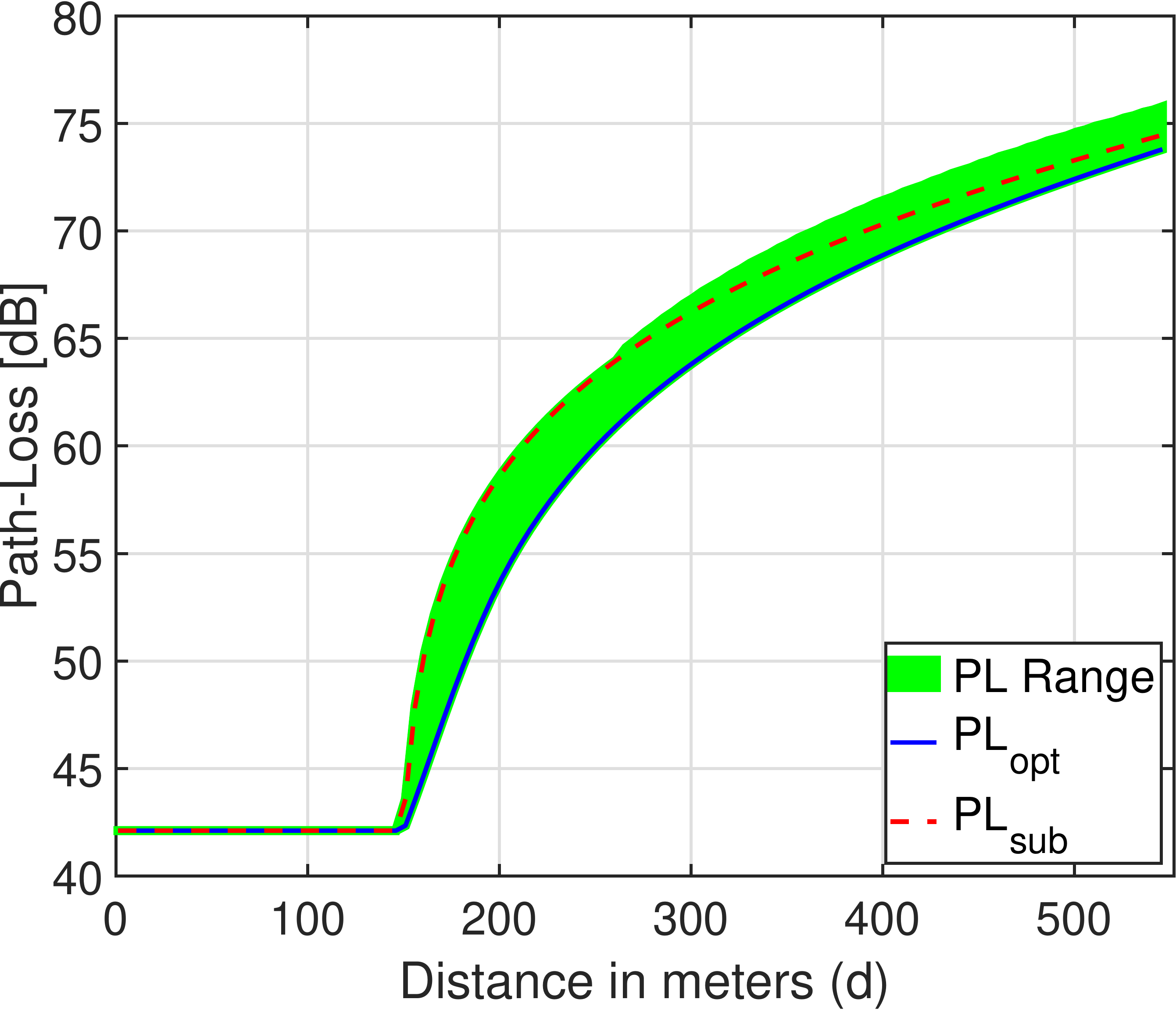}}
\end{minipage}%
\hfill
\begin{minipage}{.5\linewidth}
\centering
\subfloat[]{\label{fig:3b}\includegraphics[scale=.27]{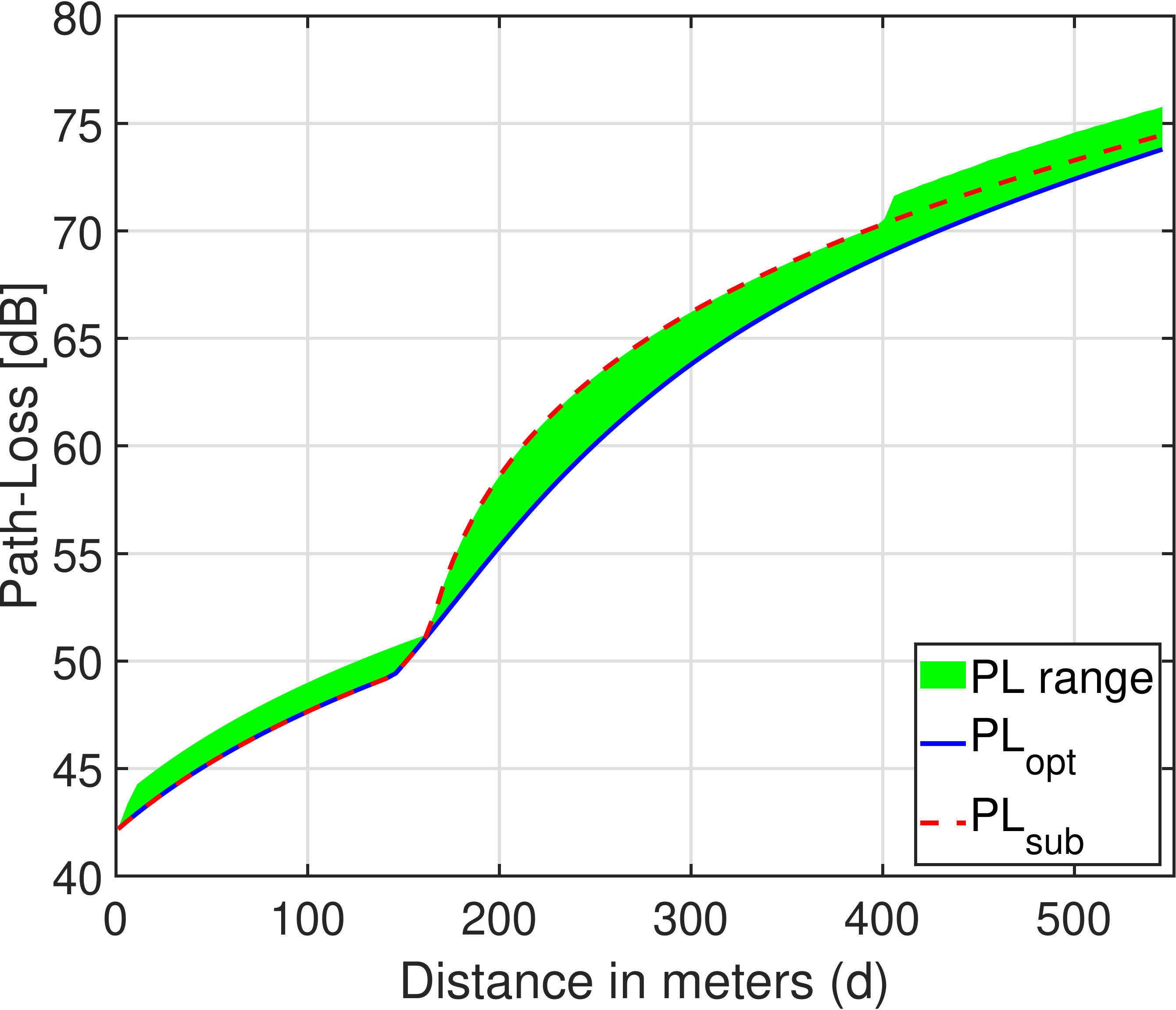}}
\end{minipage}\par\medskip
\centering
\subfloat[]{\label{fig:3c}\includegraphics[scale=.27]{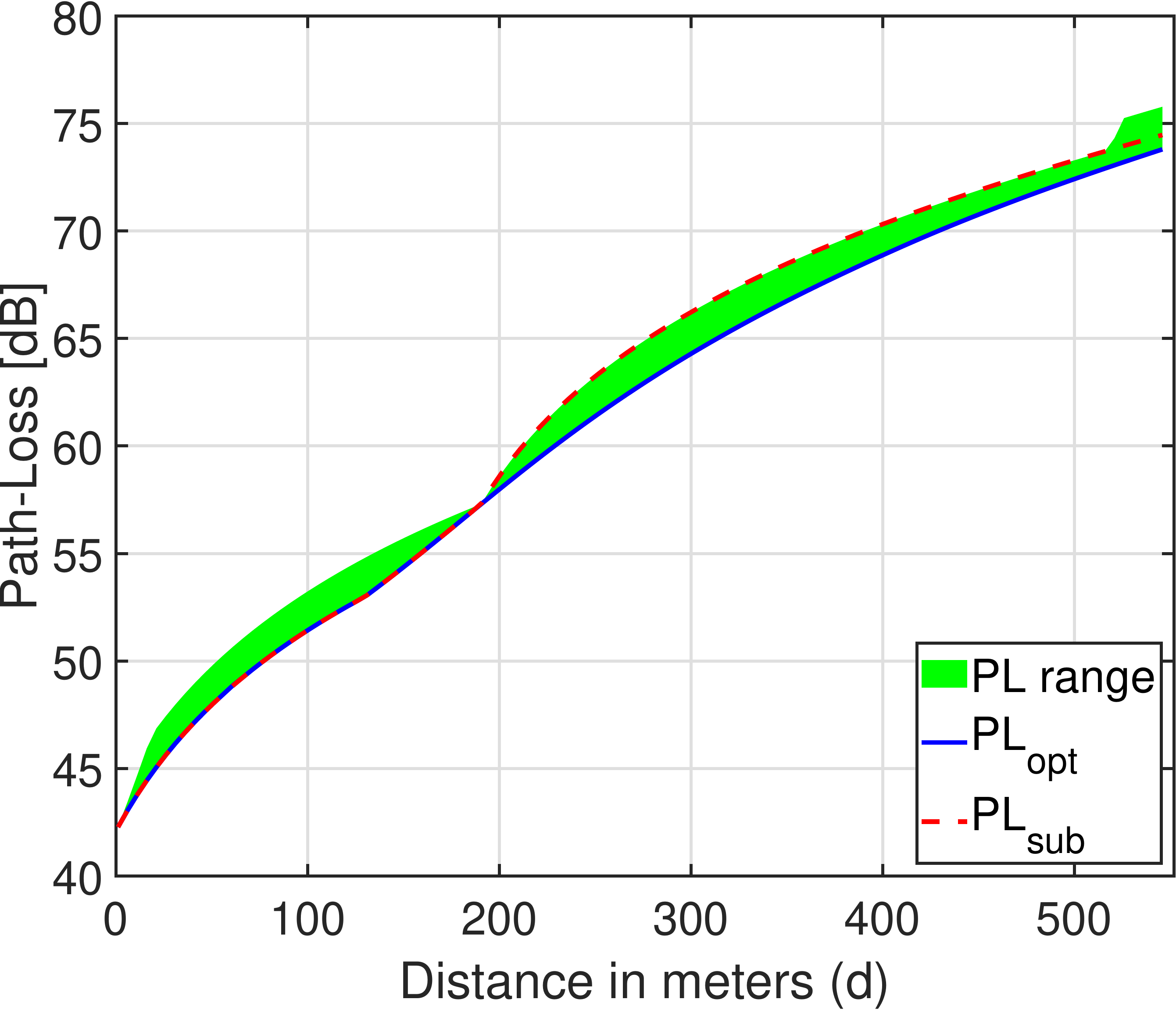}}
\caption{The PL range against different values of $d$ for: (a) $\theta_{\rm min}=0^\circ$, (b) $\theta_{\rm min}=15^\circ$, and (c) $\theta_{\rm min}=30^\circ$.}
\label{fig:main}
\end{figure}

In Fig.~\ref{fig:4}, we plot the optimal and suboptimal values of PL, for different values of $T_{\rm max}$. The results show that the gain from increasing the maximum tether length can be up to $10$ dBs reduction in the optimal value of the PL at lower values of $d$. However, as $d$ increases, the effect of increasing the value of $T_{\rm max}$ decreases. This behavior reflects some useful insights about the benefits of increasing the value of $T_{\rm max}$. When the intended users are located closer to the building, increasing the maximum tether length is very beneficial. However, if the users are located far away from the building, the effect of increasing $T_{\rm max}$ reduces. Note that value of $T_{\rm max}$ relies on lots of aspects related to the ability to control the tether tension and support its weight. That is why the value of $T_{\rm max}$ is currently limited by $150$ meters in most of the commercially available TUAV products.

As stated in Remark~\ref{rem2}, we need to evaluate how efficient are the bounds provided in Theorem~\ref{thm3} by evaluating the values of PL for all the points $p\in\mathcal{H}_{\rm opt}$. We plot this {\em PL range} for different values of $d$ with $\theta_{\rm min}=0^\circ$ in Fig.~\ref{fig:3a}, $\theta_{\rm min}=15^\circ$ in Fig.~\ref{fig:3b}, and $\theta_{\rm min}=30^\circ$ in Fig.~\ref{fig:3c}. We also plot the suboptimal value of PL, proposed in Theorem~\ref{thm4}. The results show the tightness of the PL range for all the points within the bounds provided in Theorem~\ref{thm3}. In addition, the results show that the suboptimal solution is less than $3$dBs over the optimal value for the case of $\theta_{\rm min}=15^\circ$, and it gets even tighter as $\theta_{\rm min}$ increases.
\section{Conclusion and Future Work}
Motivated by the limited lifetime and payload capacity of untethered UAVs, many companies have recently focused on developing tethered UAVs (TUAVs) with stable power supply through the tether. With the ability to achieve unlimited flight time, TUAVs unlock a whole new set of applications and use cases that were once impractical to consider for untethered UAVs due to their energy limitations. However, before considering massive deployment of TUAVs for coverage and capacity enhancement, multiple design and technical challenges should be studied carefully. This is due to the fundamental differences compared to untethered UAVs, in terms of mobility and relocation flexibility, as a result of being physically connected to the ground with a tether that has a maximum length. In this paper, we developed the first mathematical model for the deployment problem of a TUAV-enabled communication system. In particular, we considered a system of a TUAV whose launching point is placed on the rooftop of a building, and a target receiver located at distance $d$ from the building. For this setup, we first formally defined the hovering region of the TUAV, which is the set of locations achievable by the TUAV, given the tether maximum length and minimum inclination angle. Next, we used the mathematical model for the hovering region as a constraint for the optimal deployment problem. We derived upper and lower bounds for the optimal tether length and inclination angles, in order to minimize the average path-loss. In addition, we provided a closed-form expression for a suboptimal solution to the deployment problem, which maximizes the LoS probability. Using simulation results, assuming a dense urban environment, we showed the tightness of the derived upper and lower bounds, as well as the proposed suboptimal solution, in terms of average path-loss value. Finally, using tools from stochastic geometry, we derived the probability distribution of the minimum inclination angle of the tether, which highly depends on the density of the surrounding buildings and their altitudes, to avoid accidental tangling. We showed that the expected value of the minimum inclination angle increases as we move from suburban regions to high rise urban regions.

The research in the area of TUAV-enabled communication systems, in terms of analysis and design, is still taking its first steps. Hence, this work has many potential extensions. For instance, the placement of TUAVs in hotspots with known user distribution should be well investigated. In particular, given the tether length, the location of the rooftop with respect to the center of the hotspot, and the user distribution in the hotspot, the optimal tether length and inclination angle should be derived.

Given the continuous change in the user spatial distribution with time, the TUAV's optimal location actually changes with time. Hence, to study a TUAV-enabled cellular network, spatio-temporal analysis should be provided that captures the change in traffic demand both in time and space, which is another potential extension to this work.
 
\appendices
\section{Proof of Lemma~\ref{lemma1}}\label{app:lemma1}
For a given $\theta_p$, we can write $R_p^2$ as follows
\begin{align}\label{Rp2}
R_p^2=x_p^2+z_p^2&=(d-t_p\cos(\theta_p))^2+(h_b+t_p\sin(\theta_p))^2\nonumber\\&=d^2+h_b^2+t_p^2-2dt_p\cos(\theta_p)+2h_bt_p\sin(\theta_p).
\end{align}
The above function's second derivative with respect to $t_p$ is positive, hence, it is a convex function, which means that $R_p$ is also a convex function of $t_p$. Its minimum value can be found by equalizing the first derivative to zero, which leads to the final expression of $t^*$ provided in the first part of the Lemma.\\
To prove that $P_{\rm LoS}$ is an increasing function of $t_p$, we just need to show that the fraction $\frac{z_p}{x_p}$ is an increasing function of $t_p$. This is because, recalling (\ref{eqn:plos}), $P_{\rm LoS}$ is an increasing function of the fraction $\frac{z_p}{x_p}$. This fraction can be written as follows
\begin{align}
\frac{z_p}{x_p}=\frac{h_b+t_p\sin(\theta_p)}{d-t_p\cos(\theta_p)},
\end{align}  
which is clearly an increasing function of $t_p$ in the range $0<\theta_p<\frac{\pi}{2}$. This concludes the proof.
\section{Proof of Lemma~\ref{lemma2}}\label{app:lemma2}
Given that $0<\theta_p<\frac{\pi}{2}$, we can easily observe that $R_p^2$ in (\ref{Rp2}) in Appendix~\ref{app:lemma1} is an increasing function of $\theta_p$.\\
As stated in Appendix~\ref{app:lemma1}, $P_{\rm LoS}$ is an increasing function of the fraction $\frac{z_p}{x_p}$, which means that we only need to prove the concavity of $\frac{z_p}{x_p}$ as a function of $\theta_p$. Given that $\frac{z_p}{x_p}=\frac{h_b+t_p\sin(\theta_p)}{d-t_p\cos(\theta_p)}$, the first derivative is

\begin{align}\label{eqn:firstd}
\frac{t_p\cos(\theta_p)(d-t_p\cos(\theta_p))-t_p\sin(\theta_p)(h_b+t_p\sin(\theta_p))}{(d-t_p\cos(\theta_p))^2}=\frac{t_p(d\cos(\theta_p)-h_b\sin(\theta_p)-t_p)}{(d-t_p\cos(\theta_p))^2}.
\end{align}
Clearly, the first derivative in (\ref{eqn:firstd}) is a decreasing function of $\theta_p$ in the range $0<\theta_p<\frac{\pi}{2}$, which implies the concavity of $\frac{z_p}{x_p}$. The value of $\theta_p$ that maximizes $\frac{z_p}{x_p}$ can be found by equalizing the first derivative to zero, which leads to $\theta^*$ as defined in Lemma~\ref{lemma2}.
\section{Proof of Theorem~\ref{thm5}}\label{app:theta}
\begin{figure}
    \centering
\includegraphics[width=0.4\columnwidth]{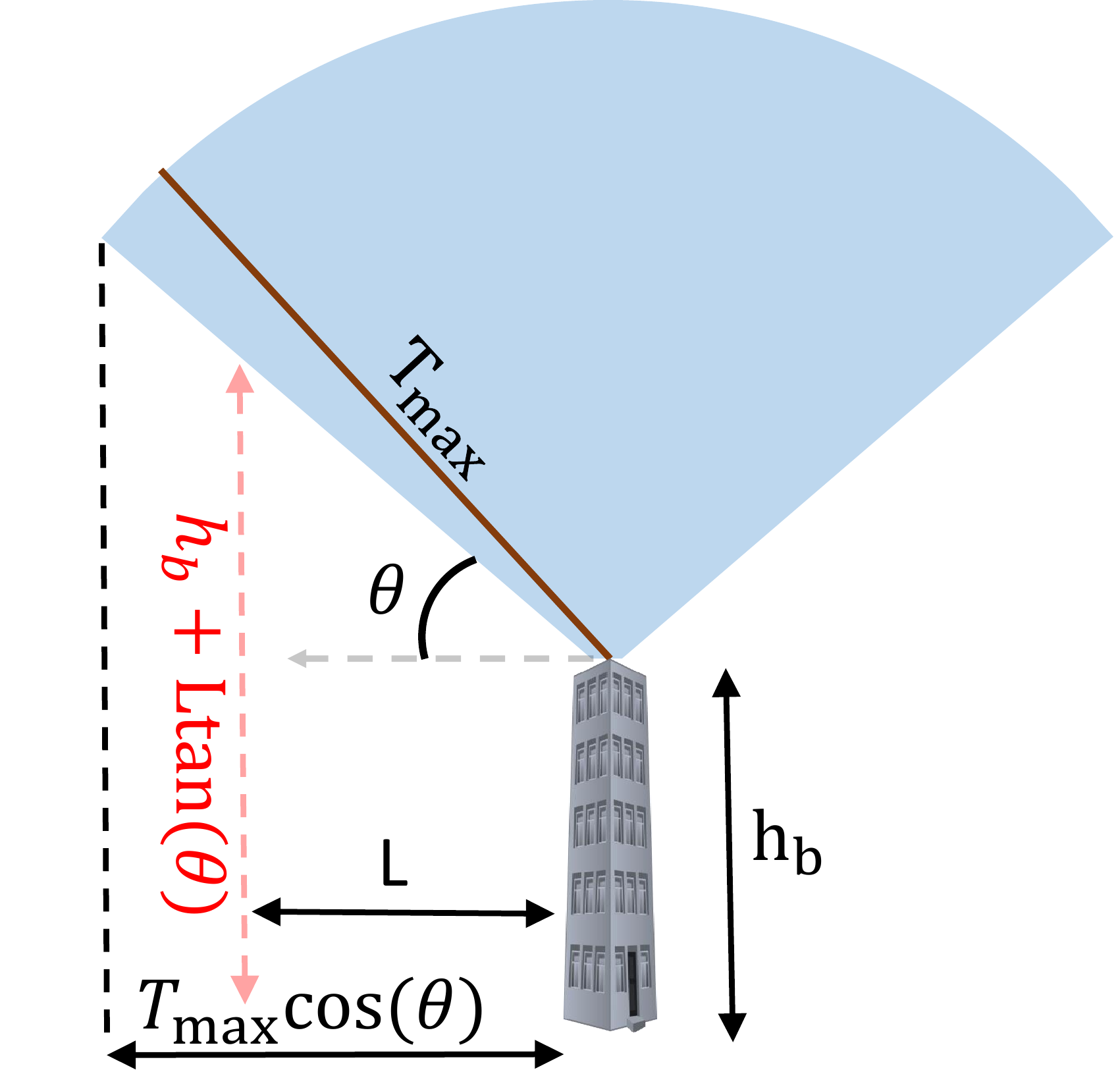}
\caption{For a given value of $\theta$, only the heights of the buildings inside the ball $\mathcal{B}(0,T_{\rm max}\cos(\theta))$ affect the value of $\mathbb{P}(\theta_{\rm min}\leq\theta)$.}
\label{fig:22}
\end{figure}
As shown in Fig.~\ref{fig:22}, assuming that the TUAV's rooftop is located at the origin, in order to ensure that $\theta_{\rm min}<\theta$, we need to ensure that any building at distance $L$ from the origin, inside the ball $\mathcal{B}(0,T_{\rm max}\cos(\theta))$, has a height $h_L<h_b+L\tan(\theta)$. Hence the CDF of $\theta_{\rm min}$ can be derived as follows
\begin{align}
\mathbb{P}(\theta_{\rm min}\leq\theta)&=\mathbb{P}\left(\underset{x_i\in\Phi_b\cap \mathcal{B}(0,T_{\rm max}\cos(\theta))}{\bigcap} h_{x_i}<h_b+|x_i|\tan(\theta)\right),
\end{align}
where $h_{x_i}$ is the height of the building located at $x_i$. Given that the set of heights $\{h_{x_i}\}$ are i.i.d with Rayleigh distribution and mean $\gamma$ then
\begin{align}\label{eqn:min}
\mathbb{P}(\theta_{\rm min}\leq\theta)&=\mathbb{E}_{\Phi_b}\left[\prod_{x_i\in\Phi_b\cap \mathcal{B}(0,T_{\rm max}\cos(\theta))}\mathbb{P}\left(h_{x_i}<h_b+|x_i|\tan(\theta)\right)\right]\nonumber\\
&=\mathbb{E}_{\Phi_b}\left[\prod_{x_i\in\Phi_b\cap \mathcal{B}(0,T_{\rm max}\cos(\theta))}\left(1-\exp\left(\frac{-(h_b+|x_i|\tan(\theta))^2}{\gamma^2}\right)\right)\right]\nonumber\\
&\overset{(a)}{=}\exp\left(-2\pi\beta\int_0^{T_{\rm max}\cos(\theta)}\exp\left(\frac{-(h_b+r\tan(\theta))^2}{\gamma^2}\right)r{\rm d}r\right),
\end{align}
where $(a)$ results from using the PGFL of PPP~\cite{haenggi2012stochastic}. Applying simple algebraic manipulations to the expression in (\ref{eqn:min}) leads to the final result in Theorem~\ref{thm5}.
\bibliographystyle{IEEEtran}
\bibliography{Draft_v0.4.bbl}

\begin{thebibliography}{10}
\providecommand{\url}[1]{#1}
\csname url@samestyle\endcsname
\providecommand{\newblock}{\relax}
\providecommand{\bibinfo}[2]{#2}
\providecommand{\BIBentrySTDinterwordspacing}{\spaceskip=0pt\relax}
\providecommand{\BIBentryALTinterwordstretchfactor}{4}
\providecommand{\BIBentryALTinterwordspacing}{\spaceskip=\fontdimen2\font plus
\BIBentryALTinterwordstretchfactor\fontdimen3\font minus
  \fontdimen4\font\relax}
\providecommand{\BIBforeignlanguage}[2]{{%
\expandafter\ifx\csname l@#1\endcsname\relax
\typeout{** WARNING: IEEEtran.bst: No hyphenation pattern has been}%
\typeout{** loaded for the language `#1'. Using the pattern for}%
\typeout{** the default language instead.}%
\else
\language=\csname l@#1\endcsname
\fi
#2}}
\providecommand{\BIBdecl}{\relax}
\BIBdecl

\bibitem{8660516}
M.~{Mozaffari}, W.~{Saad}, M.~{Bennis}, Y.~{Nam}, and M.~{Debbah}, ``A tutorial
  on {UAVs} for wireless networks: {A}pplications, challenges, and open
  problems,'' {\em IEEE Communications Surveys Tutorials}, to appear.

\bibitem{8675384}
A.~{Fotouhi}, H.~{Qiang}, M.~{Ding}, M.~{Hassan}, L.~G. {Giordano},
  A.~{Garcia-Rodriguez}, and J.~{Yuan}, ``Survey on {UAV} cellular
  communications: {P}ractical aspects, standardization advancements,
  regulation, and security challenges,'' {\em IEEE Communications Surveys
  Tutorials}, to appear.

\bibitem{8438489}
X.~{Cao}, P.~{Yang}, M.~{Alzenad}, X.~{Xi}, D.~{Wu}, and H.~{Yanikomeroglu},
  ``Airborne communication networks: A survey,'' \emph{IEEE Journal on Selected
  Areas in Communications}, vol.~36, no.~9, pp. 1907--1926, Sep. 2018.

\bibitem{zeng2019accessing}
Y.~Zeng, Q.~Wu, and R.~Zhang, ``Accessing from the sky: {A} tutorial on {UAV}
  communications for {5G} and beyond,'' 2019, available online:
  arxiv.org/abs/1903.05289.

\bibitem{8579209}
B.~{Li}, Z.~{Fei}, and Y.~{Zhang}, ``{UAV} communications for {5G} and beyond:
  {R}ecent advances and future trends,'' \emph{IEEE Internet of Things
  Journal}, vol.~6, no.~2, pp. 2241--2263, April 2019.

\bibitem{muruganathan2018overview}
S.~D. Muruganathan, X.~Lin, H.-L. Maattanen, Z.~Zou, W.~A. Hapsari, and
  S.~Yasukawa, ``An overview of {3GPP} release-15 study on enhanced {LTE}
  support for connected drones,'' 2018, available online:
  arxiv.org/abs/1805.00826.

\bibitem{8692749}
M.~M. {Azari}, F.~{Rosas}, and S.~{Pollin}, ``Cellular connectivity for {UAVs}:
  {N}etwork modeling, performance analysis and design guidelines,'' {IEEE
  Transactions on Wireless Communications}, to appear.

\bibitem{8470897}
Y.~{Zeng}, J.~{Lyu}, and R.~{Zhang}, ``Cellular-connected {UAV}: {P}otential,
  challenges, and promising technologies,'' \emph{IEEE Wireless
  Communications}, vol.~26, no.~1, pp. 120--127, Feb. 2019.

\bibitem{azariarxiv}
M.~M. Azari, G.~Geraci, A.~Garcia-Rodriguez, and S.~Pollin, ``Cellular
  {UAV}-to-{UAV} communications,'' available online: arxiv.org/abs/1904.05104.

\bibitem{8337920}
X.~{Lin}, V.~{Yajnanarayana}, S.~D. {Muruganathan}, S.~{Gao}, H.~{Asplund},
  H.~{Maattanen}, M.~{Bergstrom}, S.~{Euler}, and Y.~.~E. {Wang}, ``The sky is
  not the limit: {LTE} for unmanned aerial vehicles,'' \emph{IEEE
  Communications Magazine}, vol.~56, no.~4, pp. 204--210, April 2018.

\bibitem{8528463}
G.~{Geraci}, A.~{Garcia-Rodriguez}, L.~{Galati Giordano},
  D.~{L{\'o}pez-P{\'e}rez}, and E.~{Bj{\"o}rnson}, ``Understanding {UAV}
  cellular communications: {F}rom existing networks to massive {MIMO},''
  \emph{IEEE Access}, vol.~6, pp. 67\,853--67\,865, Nov. 2018.

\bibitem{8437232}
H.~{Wu}, X.~{Tao}, N.~{Zhang}, and X.~{Shen}, ``Cooperative {UAV}
  cluster-assisted terrestrial cellular networks for ubiquitous coverage,''
  \emph{IEEE Journal on Selected Areas in Communications}, vol.~36, no.~9, pp.
  2045--2058, Sep. 2018.

\bibitem{7470937}
Z.~{Xiao}, P.~{Xia}, and X.~{Xia}, ``Enabling {UAV} cellular with
  millimeter-wave communication: {P}otentials and approaches,'' \emph{IEEE
  Communications Magazine}, vol.~54, no.~5, pp. 66--73, May 2016.

\bibitem{7932923}
P.~{Yang}, X.~{Cao}, C.~{Yin}, Z.~{Xiao}, X.~{Xi}, and D.~{Wu}, ``Proactive
  drone-cell deployment: {O}verload relief for a cellular network under flash
  crowd traffic,'' \emph{IEEE Transactions on Intelligent Transportation
  Systems}, vol.~18, no.~10, pp. 2877--2892, Oct. 2017.

\bibitem{7470933}
Y.~{Zeng}, R.~{Zhang}, and T.~J. {Lim}, ``Wireless communications with unmanned
  aerial vehicles: {O}pportunities and challenges,'' \emph{IEEE Communications
  Magazine}, vol.~54, no.~5, pp. 36--42, May 2016.

\bibitem{7744808}
I.~{Bor-Yaliniz} and H.~{Yanikomeroglu}, ``The new frontier in {RAN}
  heterogeneity: {M}ulti-tier drone-cells,'' \emph{IEEE Communications
  Magazine}, vol.~54, no.~11, pp. 48--55, Nov. 2016.

\bibitem{8360023}
A.~M. {Hayajneh}, S.~A.~R. {Zaidi}, D.~C. {McLernon}, M.~{Di Renzo}, and
  M.~{Ghogho}, ``Performance analysis of {UAV} enabled disaster recovery
  networks: {A} stochastic geometric framework based on cluster processes,''
  \emph{IEEE Access}, vol.~6, pp. 26\,215--26\,230, 2018.

\bibitem{7888557}
Y.~{Zeng} and R.~{Zhang}, ``Energy-efficient {UAV} communication with
  trajectory optimization,'' \emph{IEEE Transactions on Wireless
  Communications}, vol.~16, no.~6, pp. 3747--3760, June 2017.

\bibitem{8320772}
A.~{Fotouhi}, M.~{Ding}, and M.~{Hassan}, ``Flying drone base stations for
  macro hotspots,'' \emph{IEEE Access}, vol.~6, pp. 19\,530--19\,539, March
  2018.

\bibitem{8570843}
M.~A. {Abd-Elmagid} and H.~S. {Dhillon}, ``Average peak age-of-information
  minimization in {UAV}-assisted {IoT} networks,'' \emph{IEEE Transactions on
  Vehicular Technology}, vol.~68, no.~2, pp. 2003--2008, Feb. 2019.

\bibitem{8247211}
Q.~{Wu}, Y.~{Zeng}, and R.~{Zhang}, ``Joint trajectory and communication design
  for {Multi-UAV} enabled wireless networks,'' \emph{IEEE Transactions on
  Wireless Communications}, vol.~17, no.~3, pp. 2109--2121, March 2018.

\bibitem{8053918}
M.~{Mozaffari}, W.~{Saad}, M.~{Bennis}, and M.~{Debbah}, ``Wireless
  communication using unmanned aerial vehicles {(UAVs)}: {O}ptimal transport
  theory for hover time optimization,'' \emph{IEEE Transactions on Wireless
  Communications}, vol.~16, no.~12, pp. 8052--8066, Dec. 2017.

\bibitem{7967745}
V.~V. {Chetlur} and H.~S. {Dhillon}, ``Downlink coverage analysis for a finite
  {3-D} wireless network of unmanned aerial vehicles,'' \emph{IEEE Transactions
  on Communications}, vol.~65, no.~10, pp. 4543--4558, Oct. 2017.

\bibitem{7412759}
M.~{Mozaffari}, W.~{Saad}, M.~{Bennis}, and M.~{Debbah}, ``Unmanned aerial
  vehicle with underlaid device-to-device communications: {P}erformance and
  tradeoffs,'' \emph{IEEE Transactions on Wireless Communications}, vol.~15,
  no.~6, pp. 3949--3963, June 2016.

\bibitem{8255733}
T.~{Long}, M.~{Ozger}, O.~{Cetinkaya}, and O.~B. {Akan}, ``Energy neutral
  internet of drones,'' \emph{IEEE Communications Magazine}, vol.~56, no.~1,
  pp. 22--28, Jan. 2018.

\bibitem{8648453}
B.~{Galkin}, J.~{Kibilda}, and L.~A. {DaSilva}, ``{UAVs} as mobile
  infrastructure: {A}ddressing battery lifetime,'' \emph{IEEE Communications
  Magazine}, vol.~57, no.~6, pp. 132--137, June 2019.

\bibitem{TUAVmag2019}
M.~A. Kishk, A.~Bader, and M.-S. Alouini, ``Capacity and coverage enhancement
  using long-endurance tethered airborne base stations,'' 2019, submitted to
  IEEE Communications Magazine. {A}vailable online: arxiv.org/abs/1906.11559.

\bibitem{8255764}
M.~{Alzenad}, M.~Z. {Shakir}, H.~{Yanikomeroglu}, and M.-S. {Alouini},
  ``{FSO}-based vertical backhaul/fronthaul framework for {5G+} wireless
  networks,'' \emph{IEEE Communications Magazine}, vol.~56, no.~1, pp.
  218--224, Jan. 2018.

\bibitem{7470932}
S.~{Chandrasekharan}, K.~{Gomez}, A.~{Al-Hourani}, S.~{Kandeepan},
  T.~{Rasheed}, L.~{Goratti}, L.~{Reynaud}, D.~{Grace}, I.~{Bucaille},
  T.~{Wirth}, and S.~{Allsopp}, ``Designing and implementing future aerial
  communication networks,'' \emph{IEEE Communications Magazine}, vol.~54,
  no.~5, pp. 26--34, May 2016.

\bibitem{cicek2018backhaul}
C.~T. Cicek, T.~Kutlu, H.~Gultekin, B.~Tavli, and H.~Yanikomeroglu,
  ``Backhaul-aware placement of a {UAV-BS} with bandwidth allocation for
  user-centric operation and profit maximization,'' 2018, available online:
  arxiv.org/abs/1810.12395.

\bibitem{8422376}
B.~{Galkin}, J.~{Kibilda}, and L.~A. {DaSilva}, ``Backhaul for low-altitude
  {UAVs} in urban environments,'' in \emph{in Proc., International Conference
  on Communications (ICC)}, May 2018, pp. 1--6.

\bibitem{equinox}
Equinox Innovative Systems, ``DELTA {3C}: Tri-Sector Cell Tower'', 2017.
  Available online: https://t2m.io/UaF8ZYbq. [accessed on July 2nd, 2019].

\bibitem{tds}
Tethered Drone Systems, ``Tethered Drone Systems: The Future of Tethered {UAV}
  Technology'', 2019. Available online: https://t2m.io/eMsFWd5P. [accessed on
  July 2nd, 2019].

\bibitem{aria}
Aria Insights, ``PARC: The Future of High-powered Commercial Drones''.
  Available online: https://t2m.io/bufh6WRy. [accessed on July 2nd, 2019].

\bibitem{elistair}
Elistair, ``Orion: Persistent {UAV} for Surveillence and Communications'',
  2014. Available online: https://t2m.io/5LeDMh9S. [accessed on July 2nd,
  2019].

\bibitem{abs}
European Union FP7, ``{ABSOLUTE} - Aerial Base Stations with Opportunistic
  Links for Unexpected {$\&$} Temporary Events'', 2019. Available online:
  https://t2m.io/RbeVUhSr. [accessed on July 2nd, 2019].

\bibitem{sundaresan2018skylite}
K.~Sundaresan, E.~Chai, A.~Chakraborty, and S.~Rangarajan, ``Sky{L}ite:
  {E}nd-to-end design of low-altitude {UAV} networks for providing {LTE}
  connectivity,'' 2018, available online: arxiv.org/abs/1802.06042.

\bibitem{8644135}
M.~Y. {Selim} and A.~E. {Kamal}, ``Post-disaster {4G/5G} network rehabilitation
  using drones: {S}olving battery and backhaul issues,'' in \emph{In Proc.,
  IEEE Globecom Workshops}, Dec 2018, pp. 1--6.

\bibitem{8432474}
Q.~{Wu}, J.~{Xu}, and R.~{Zhang}, ``Capacity characterization of {UAV}-enabled
  two-user broadcast channel,'' \emph{IEEE Journal on Selected Areas in
  Communications}, vol.~36, no.~9, pp. 1955--1971, Sep. 2018.

\bibitem{6863654}
A.~{Al-Hourani}, S.~{Kandeepan}, and S.~{Lardner}, ``Optimal {LAP} altitude for
  maximum coverage,'' \emph{IEEE Wireless Communications Letters}, vol.~3,
  no.~6, pp. 569--572, Dec. 2014.

\bibitem{7918510}
M.~{Alzenad}, A.~{El-Keyi}, F.~{Lagum}, and H.~{Yanikomeroglu}, ``{3-D}
  placement of an unmanned aerial vehicle base station {(UAV-BS)} for
  energy-efficient maximal coverage,'' \emph{IEEE Wireless Communications
  Letters}, vol.~6, no.~4, pp. 434--437, Aug. 2017.

\bibitem{7762053}
J.~{Lyu}, Y.~{Zeng}, R.~{Zhang}, and T.~J. {Lim}, ``Placement optimization of
  {UAV}-mounted mobile base stations,'' \emph{IEEE Communications Letters},
  vol.~21, no.~3, pp. 604--607, March 2017.

\bibitem{7486987}
M.~{Mozaffari}, W.~{Saad}, M.~{Bennis}, and M.~{Debbah}, ``Efficient deployment
  of multiple unmanned aerial vehicles for optimal wireless coverage,''
  \emph{IEEE Communications Letters}, vol.~20, no.~8, pp. 1647--1650, Aug.
  2016.

\bibitem{7510820}
R.~I. {Bor-Yaliniz}, A.~{El-Keyi}, and H.~{Yanikomeroglu}, ``Efficient {3-D}
  placement of an aerial base station in next generation cellular networks,''
  in \emph{In Proc., International Conference on Communications (ICC)}, May
  2016, pp. 1--5.

\bibitem{8038869}
M.~{Mozaffari}, W.~{Saad}, M.~{Bennis}, and M.~{Debbah}, ``Mobile unmanned
  aerial vehicles {(UAVs)} for energy-efficient internet of things
  communications,'' \emph{IEEE Transactions on Wireless Communications},
  vol.~16, no.~11, pp. 7574--7589, Nov. 2017.

\bibitem{7037248}
A.~{Al-Hourani}, S.~{Kandeepan}, and A.~{Jamalipour}, ``Modeling air-to-ground
  path loss for low altitude platforms in urban environments,'' in \emph{in
  Proc., Global Communications Conference}, Dec. 2014, pp. 2898--2904.

\bibitem{haenggi2012stochastic}
M.~Haenggi, \emph{Stochastic {G}eometry for {W}ireless {N}etworks}.\hskip 1em
  plus 0.5em minus 0.4em\relax Cambridge University Press, 2012.

\end{thebibliography}
\end{document}